\documentclass[11pt]{article}

\usepackage{subfigure}
\usepackage{latexsym, mathrsfs}
\usepackage{amsfonts}
\usepackage{amsmath, amsthm, amssymb}
\usepackage{epsfig}
\usepackage{color}
\usepackage{wrapfig}
\usepackage{graphicx}
\usepackage[sort&compress]{natbib}
\bibpunct{[}{]}{,}{n}{,}{,}

\usepackage[T1]{fontenc}
\usepackage{bezier}

\usepackage{setspace}
\begin{document}

\newcommand{\no}{\noindent}
\newcommand{\be}{\begin{eqnarray}}
\newcommand{\ee}{\end{eqnarray}}
\newcommand{\beeq}{\begin{equation}}
\newcommand{\eeeq}{\end{equation}}
\newcommand{\beqs}{\begin{eqnarray*}}
\newcommand{\eeqs}{\end{eqnarray*}}
\newcommand{\bms}{\boldsymbol}

\newtheorem{theorem}{\bf Theorem}
\newtheorem{remark}{\bf Remark}
\newtheorem{result}{Result}
\newtheorem{observation}{\bf Observation}
\newtheorem{corollary}{\bf Corollary}
\newtheorem{definition}{Definition}
\newtheorem{lemma}{\bf Lemma}
\newtheorem{proposition}{\bf Proposition}

\title{
Energy Efficient Estimation of  Gaussian Sources Over Inhomogeneous Gaussian MAC Channels 
}

\date{}
\author{Shuangqing Wei, Rajgopal Kannan, Sitharama Iyengar and  Nageswara S.
Rao}

\maketitle 
\footnotetext[1]{S. Wei is with the Department of ECE, Louisiana State
University, Baton Rouge} \footnotetext[2]{R. Kannan and S. Iyengar are with the Department of CS,
Louisiana State University, Baton Rouge} \footnotetext[3]{N. Rao is with the Computer
Science and Mathematics Division, Oak Ridge National Laboratory}
\footnotetext[4]{This work is funded in part from DOE-ORNL
(Sensornets Program Sept. 2006- 2008)}

\begin{abstract}
It has been shown lately
the optimality of uncoded transmission in
estimating Gaussian sources over homogeneous/symmetric Gaussian multiple
access channels (MAC) using multiple sensors.
It remains, however,  unclear whether it still holds for any arbitrary networks and/or
with high channel signal-to-noise ratio (SNR)  and
high signal-to-measurement-noise ratio (SMNR). In this paper, we first
provide a  joint source and channel coding approach in estimating
Gaussian sources over Gaussian MAC channels, as well as its sufficient and
necessary condition in restoring Gaussian sources with a prescribed
distortion value. An interesting relationship between our proposed joint
approach with a more straightforward separate source and channel coding
scheme is then established. Further comparison of these two schemes with the
uncoded approach reveals a lacking of a  consistent ordering of these three
strategies in terms of the total transmission power consumption under a
distortion constraint for arbitrary in-homogeneous networks. We then
formulate constrained power minimization problems and transform them to relaxed
convex geometric programming problems, whose numerical results exhibit
that  either separate or  uncoded scheme becomes dominant over a linear topology
network. In addition, we prove that the optimal decoding order to minimize
the total transmission powers for both
source and channel coding parts is solely
subject to the ranking of MAC channel qualities, and has nothing to do with
the ranking of measurement qualities.
Finally, asymptotic results for homogeneous networks are obtained which not only confirm
the existing optimality of the uncoded approach, but also show that the
asymptotic SNR exponents of these three approaches are all the same.
Moreover, the proposed joint approach share the same asymptotic  ratio with
respect to high SNR and high SMNR as the uncoded scheme.
\end{abstract}

\section{Introduction}
\label{sec:intro}

Recent years have witnessed a tremendous growth of interests in
wireless ad hoc and sensor networks from both academia and
industry, due to their ease of implementation, infrastructure-less
nature, as well as the huge potentials in civil and military
applications. In many instances, sensor nodes  are deployed to
serve a common purpose such as  surveillance  and monitoring
environments. One of the major design metrics is to maximize the
lifetime of a sensor net while meeting the constraints imposed by
the quality of reconstruction, such as the resulting ultimate
distortion measure when data collected by sensors are fused  to
construct an estimate of  the monitored source. A critical factor
affecting the lifetime of a sensor net is the amount of total
power expenditure that  senor nodes spend on transmitting their
measurements to a fusion center. The power consumptions are
closely related with the way sensors collecting and processing
measurements, as well as the communication link quality between
sensor nodes and the fusion center.

In this paper, assuming $L$ sensor nodes send measurements of a
Gaussian source to a fusion center via a one-hop interference
limited wireless link, we investigate the issue of power
allocations across sensors with and without local compression and channel
coding.
A similar
system model for Gaussian sensor networks has  also been
adopted recently by  \cite{gastpar_IT05, gastpar_JSAC_05, Gastpar_exact_optimality} and
\cite{behroozi_ICASSP06}. In \cite{gastpar_IT05, gastpar_JSAC_05},
the authors investigated joint source-channel coding paradigm and
analyzed how distortion scales with respect to the size of the
network. They showed that uncoded transmission achieves the lower
bound of the mean squared error distortion as the number of
sensors grow to infinity in  {\em symmetric} networks.
However, no
exact source and channel coding schemes are provided for general
system settings other than the uncoded  scheme.
In  \cite{Gastpar_exact_optimality}, the ``exact"  optimality of uncoded transmission is
proved even for the homogeneous Gaussian networks with {\em finite} number of sensor nodes.
As pointed out in \cite{Gastpar_exact_optimality}, it  remains
unclear though what approach is more favorable when a system
becomes non-symmetric with a finite number of sensors.

The objectives of this paper are two folds. First, we will propose a
joint source-channel coding approach and then establish its relationship
with the separate source and channel coding strategy. Second, we will
investigate the optimal rate and power allocation strategy in order to
minimize  the total transmission power  under the constraint that the
mean squared error value in estimating the Gaussian source remotely is no
greater than a prescribed threshold.
In particular, we will compare the resulting total power consumptions of
three distinct processing schemes, namely, joint source and channel coding,
separate source and channel coding and uncoded amplify-and-forward
approaches for in-homogeneous networks, and demonstrate the well known
result of the optimality of uncoded approach for estimating Gaussian
sources in homogeneous networks does not always hold in inhomogeneous networks.

Our contributions in this paper can be summarized as follows:
\begin{itemize}

\item A  joint source and channel coding approach is proposed, whose
achievable rate region is obtained. An interesting relationship between
this approach and separate source and channel coding approach is then
established.

\item Optimal decoding order for both joint and separate source channel
coding is found which is only a function of MAC channel ranking order, and
has nothing to do with the power level of  source measurement noise, when we
intend to minimize the total transmission power.

\item Relaxed geometric programming problems are formulated in order to
compare three schemes. Numerical results demonstrate the uncoded
transmission is not always the best option. The ordering of the three
schemes is highly dependent on relative channel qualities, measurement
noise levels, as well as the distortion threshold.

\item  Asymptotic results for large size homogeneous networks with finite
SNR and SMNR are obtained which show the optimality of uncoded transmission
from another perspective. More importantly, a condition is found under
which the scaling factor of received channel SNR versus signal-to-distortion-noise-ratio
(SDNR),  as SMNR  grows to infinity,  of joint approach is equal to that
of the  uncoded scheme. In addition, we prove the SNR exponents of all three schemes
are the same.

\end{itemize}

The paper is organized as follows. System model is set up in
Section~\ref{model}. A joint source and channel coding scheme is
proposed in Section~\ref{joint}, in which we establish its
achievable rate region, as well as its relationship with the
separate source and channel coding scheme. In order to compare
joint and separate approaches, the formulated total power
minimization problems are solved using geometric programming
approach in Section~\ref{optimization}, where we also obtain the
optimal decoding order for  non-homogeneous networks. Uncoded
approach is revisited in Section~\ref{uncoded} from the
perspective of making comparisons with the former two approaches.
In Section~\ref{asymptotic comparision}, we compare the
aforementioned three schemes in asymptotic region for homogeneous
networks. Finally, numerical results of our comparisons for
arbitrary two-node networks are presented in
Section~\ref{simulation}.

\section{System Model} \label{model}

\begin{figure}[ht]
         \centering
\centerline{
\scalebox{0.8}{
\input{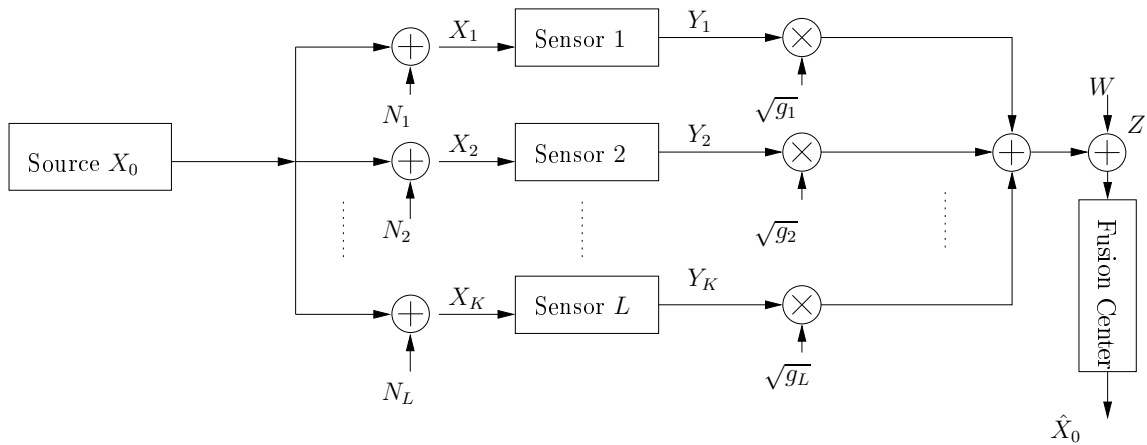}
}
}
\caption{System model }
\end{figure}

Assume $L$ sensor nodes observe a common Gaussian source $X_0[i],
i = 1, \cdots, n$, where  $X_0[i] \sim \mathcal{N}(0, \sigma_S^2)$
are identically and independently distributed Gaussian random
variables with mean zero and variance $\sigma^2_S$. The
measurements $X_j[i] = X_0[i] + N_j[i]$, $j=1, \cdots, L$ from $L$
sensors experience independent additive Gaussian measurement noise
$N_j[i] \sim \mathcal{N}(0, \sigma_{N_j}^2)$, where independence
is assumed to hold  across both space and time. Let $Y_j[i]$
denote the transmitted signal from sensor $j$ at time $i$, which
satisfies  an average power constraint: \beeq
\frac{1}{n}\sum_{i=1}^n  \left| Y_j[i] \right|^2 \leq P_j,
\, j=1 \cdots, L. \eeeq.

The processed signals $\{Y_j[i]\}$ then go through a Gaussian multiple
access  channel  and are  superposed at a fusion center
resulting in $Z[i] = \sum_{j=1}^L \sqrt{g_j}Y_j[i] + W[i]$, where
$W[i] \sim \mathcal{N}(0, \sigma_W^2)$ are  white Gaussian noise
introduced at the fusion center and assumed independent with
$N_j[i]$.
  Coefficients  $g_j,
j=1, \cdots, L$ capture the underlying channel pathloss and fading
from sensors to the fusion center. In this paper, we assume
coherent fusion is conducted  in the sense that  $g_j$ are assumed
perfectly known by the fusion
center.  Upon receiving $\{Z[i]\}$, the fusion center constructs an estimate
$\{\hat{X}_0[i]\}$ of $\{X_0[i]\}$ such that the average mean
squared error $D_E \stackrel{\Delta}{=} \lim_{n \rightarrow
\infty} \frac{1}{n} \sum_{i=1}^n E \left|X_0[i] -
\hat{X}_0[i]\right|^2$ of the estimation  satisfies $D_E \leq D$,
where $D$ is a prescribed upper bound for estimation error.

What interests us in this paper is power efficient schemes to estimate the Gaussian
source remotely with a prescribed mean squared error. Three approaches, namely, joint source
and channel coding, separate source and channel coding, and uncoded amplify-and-forward
schemes,  will be investigated in the sequel.
\section{Joint Source-Channel Based Fusion and Its Relationship with Separate Source and Channel Coding}
\label{joint}

In \cite{lapidoth_ISIT06_1}, a joint source and channel coding scheme is
proposed to estimate  two correlated Gaussian sources remotely at a fusion center where
measurements from two sensors are received through a Gaussian MAC channel.  Achievable rate
region was obtained as a function of the required distortion tuple in restoring two correlated
sources.  Inspired by their work, we,
in this section, will first develop an achievable rate region for our
proposed joint source-channel coding (JSCC) approach for any arbitrary network with
$L>1$ sensor nodes and then demonstrate an interesting
relationship of JSCC with a separate source and channel coding scheme
(SSCC) which is a straightforward combination of the recent findings on
CEO problem  \cite{oohama_IT05}
and traditional MAC channel coding \cite{cover_book} with independent
sources.

\subsection{Achievable Rate Region for Distributed Joint Source-Channel Coding
in Estimation of  Gaussian Sources}
\label{joint region}

Let $\tilde{R}_j, j=1,\cdots, L$ denote the
compression rate at the $j$-th sensor. There are total $2^{n \tilde{R}_j}$
source codewords ${\bf U}_j = \{ {\bf U}^{(k)}_j,  k= 1, \cdots, 2^{n \tilde{R_j}}\}$ from which 
sensor $j$ selects   ${\bf U}^{(m_j)}_j = \{  U^{(m_j)}_j[i], i=1,\cdots, n \}$ 
to  represent ${\bf X_j}= \{ X_j[i], i =1, \cdots, n\}$.
 The joint approach we propose here is to let
each sensor directly transmit a scaled version of a source
codeword ${\bf U}^{(m_j)}_j $. The scaling factor introduced herein
is to maintain the average transmission power $P_j$ by sensor $j$,
$j=1, \cdots, L$. Since $L$ sensors see the same Gaussian source
with independent measurement noise, $L$ quantization vectors
$\{{\bf U}_j^{(m_j)}, j=1, \cdots, L\}$ are correlated. As a result, the
decoding at fusion center needs to take into account of such
correlation when it performs joint decoding of these $L$
codewords. The decoded source/channel codeword ${\bf \hat{U}}^{(m_j)}_j $
are then linearly combined to obtain an MMSE estimate
$\{\hat{X}_0[i], i=1, \cdots, n\}$ of the Gaussian source
$\{X_0[i], i=1, \cdots, n\}$.

We are interested in deriving the achievable region of rate tuples
$\{\tilde{R}_j, j=1, \cdots, L\}$ such that $2^{n \tilde{R}_j}, j=1,\cdots, L$
source/channel codewords can be decoded with asymptotic zero error
and the mean squared error $D_E$ satisfies $D_E \leq D$.

\begin{theorem} \label{theorem2}

To make $D_E \leq D$, $\tilde{R}_i$ satisfy
\beeq \label{theorem2:0}
\tilde{R}_i = I(X_i; U_i) = r_i + \frac{1}{2} \log \left[ 1 +
\frac{\sigma_S^2}{\sigma^2_{N_i}}\left(1 -2^{-2r_i}\right) \right],\; i =1,
\cdots, L
\eeeq

\no where $r_i \geq 0, i=1, \cdots, L$ are chosen based on
\beeq \label{MSE L}
\frac{1}{D_E} = \frac{1}{\sigma_S^2} + \sum_{k=1}^{L}
\frac{1-2^{-2r_k}}{\sigma_{N_k}^2} \geq \frac{1}{D}
\eeeq

\no and $I(X_j;U_j)$ denotes the  mutual information between $X_j$ and a Gaussian
random variable $U_j$, which is associated with $X_j$ by
\beeq \label{theorem2:1}
U_j = X_j + V_j, j =1, \cdots, L
\eeeq

\no where $V_j$, independent of $X_j$, are  independent Gaussian random
variables with mean $0$ and variance $\sigma^2_{V_i} =
\sigma^2_{N_i}/(2^{r_i}-1)$.
\end{theorem}
\begin{proof}
The proof is a straightforward application of the techniques used
in proving Lemma~10 in \cite{oohama_IT05}. For brevity, we only
provide an outline here.

We quantize $\{X_j[i]\}$ with $2^{n \tilde{R}_j}$ Gaussian vectors
$\{\hat{U}_j[i]\}$ such that the source symbol $X_j [i]$ can be
constructed from the quantized symbol through a test channel
\cite{cover_book}: $X_j [i] = \hat{U}_j[i] + \hat{V}_j[i]$, where
$\hat{V}_j[i]$ is a Gaussian random variable with mean zero and
variance $2^{-2 \tilde{R}_j} \sigma^2_{X_j}$, which is independent
of $\hat{U}_j[i]\sim \mathcal{N}\left(0, (1- 2^{-2 \tilde{R}_j})
\sigma_{X_j}^2\right)$. Equivalently, we can also represent
$\hat{U}_j[i]$ as $\hat{U}_j[i] = \alpha X_j [i] + \tilde{V}_j[i]$,
 where
$\alpha$ is the linear-MMSE estimate coefficient  and
$\tilde{V}_j[i]$ is the resultant estimation error. By orthogonal
principle, we have
\[\alpha = \frac{\sigma^2_{\hat{U}_j}}{\sigma^2_{X_j}} = \left(1 -
2^{-2{\tilde{R}_j}}\right)\]

\no and $\tilde{V}_j[i]$ is a Gaussian variable independent of
$X_j [i]$ with mean zero and variance $2^{-2 \tilde{R}_j} \left(1
- 2^{-2{\tilde{R}_j}}\right) \sigma^2_{X_j}$. Therefore, after
normalization, we obtain \beeq U_j = \frac{1}{\alpha} \hat{U}_j =
X_j + \frac{1}{\alpha}\tilde{V}_j = X_j + V_j \eeeq

\no where $V_j \sim \mathcal{N}\left(0, \sigma^2_{X_j}/(2^{2
\tilde{R}_j}-1)\right)$. We introduce variables $r_j$ such that
\beeq \label{R with r} 2^{2 \tilde{R}_j} -1  = \left( 2 ^{2 r_j} -1
\right) \frac{\sigma^2_{X_j}}{\sigma^2_{N_j}} \eeeq

\no which proves (\ref{theorem2:1}). We can also see that $r_j$ is actually the conditional
mutual information between $X_j$ and $U_j$ given $X_0$, i.e. $r_j = I\left(X_j; U_j
|X_0\right)$.
Since $I(X_j; U_j ) = H(U_j) - H(V_j)$,  it is then straightforward to show that
(\ref{theorem2:0}) holds. 

Given $U_j = X_j + V_j$ and $X_j = X_0 + N_j$, where $N_j$ and
$V_j$ are independent, we can construct the LMMSE estimate of
$X_0$ by $ \hat{X}_0 = \sum_{j=1}^L \beta_j U_j$,  where
coefficients $\beta_j$ can be determined again using Orthogonal
Principle.
Based on Equations (95) and (96) in \cite{oohama_IT05}, we obtain the desired
result for the mean
squared error in (\ref{MSE L}).

\end{proof}

From the proof of Theorem~\ref{theorem2}, it can be seen that $U_i$ and $U_j$ are correlated
due to the correlation between $X_i$ and $X_j$, whose correlation can be captured by
$\rho_{i,j}$, the covariance coefficient between $X_i$ and
$X_j$, which   can be computed as  
\beeq \label{direct rho} 
\rho_{i,j}  = \frac{E [X_i X_j
]}{\sqrt{E |X_i|^2 E |X_j|^2 }}=
\frac{\sigma_S^2}{\sqrt{(\sigma_S^2 + \sigma_{N_i}^2)(\sigma_S^2+
\sigma_{N_j}^2)}} \eeeq

\no The covariance coefficient $\tilde{\rho}_{i,j}$ between $U_i$ and
$U_j$ can be obtained accordingly as: \beeq \label{two rhos}
 \tilde{\rho}_{i,j}= \rho_{i,j}
\sqrt{(1- 2^{-2 \tilde{R}_i})(1- 2^{-2 \tilde{R}_j})}. \eeeq

\no  After substituting $\tilde{R}_i$ determined in
Theorem~\ref{theorem2} into it, we obtain \beeq \label{new rho}
\tilde{\rho}_{i,j} = \sqrt{\frac{q_i q_j}{(1+q_i)(1+q_j)}}, \eeeq

\no where $q_i = \frac{\sigma_S^2}{\sigma_{N_i}^2} (1 - 2^{-2 r_i})$.

\no For any given subset $S \subseteq \{1, 2, \cdots, L\}$, define 
vectors ${\bf U(S)} = \left[U_{\pi_1}, \cdots, U_{\pi_{|S|}} \right]$ and 
${\bf U(S^c)} = \left[U_{\pi_{|S|+1}}, \cdots, U_{\pi_L} \right]$, where $\pi$ is an
arbitrary ordering of the $L$ indexes. The  
covariance matrix  of  ${\bf U} = [{\bf U(S)}, {\bf
U(S^c)}]^T$ can thus  be decomposed as
\beeq \label{cov_matrix}
{\bf \Sigma_{U}} = E \left[ {\bf U} {\bf U}^T\right] =  \left[
\begin{array}{ll} {\bf \Sigma_S} & {\bf \Sigma_{S,S^c}}\\
{\bf \Sigma_{S^c,S}} & {\bf \Sigma_{S^c}} 
\end{array} \right],  \eeeq

\no where  ${\bf \Sigma_S}$, ${\bf
\Sigma_{S^c}}$, ${\bf \Sigma_{S,S^c}}$ denote 
the auto- and cross-covariance matrices of ${\bf U(S)}$ and ${\bf U(S^c)}$. 
The entries of ${\bf \Sigma_{U}}$ are $\left({\bf
\Sigma_{U}}\right)_{i,j} = \tilde{\rho}_{i,j} \sqrt{P_i P_j}$  for
$i \neq j$ and  $\left({\bf \Sigma_{U}}\right)_{i,i} =P_i$, $i, j
\in \{1,\cdots, L\}$, where $\tilde{\rho}_{i,j}$ is obtained in (\ref{new rho}).

After each sensor maps the observation vector to $U_j$, an additional scaling factor
$\gamma_j = \sqrt{\frac{P_j}{\sigma_{U_j}^2}}$ is imposed on $U_j$, where $\sigma_{U_j}^2 =
\sigma_{X_j}^2/\left(1-2^{-2 \tilde{R}_j} \right)$ in order to keep the average transmission
power of $Y_j[i] = \gamma_j U_j[i] $ as $P_j$. The received signal at the fusion center can thus be written as
\beeq \label{RX fusion}
Z[i]= \sum_{j=1}^L \gamma_j U_j[i] \sqrt{g_j} + W[i]
\eeeq

\begin{theorem} \label{theorem correlated}
Given the received signal $Z[i], i=1, \cdots, n $ in (\ref{RX fusion}), the quantization rate $\tilde{R}_i,
i=1, \cdots, L$ obtained  in Theorem~\ref{theorem2} satisfies the following inequalities in order
to restore $X_0$ at the fusion center with distortion no less than $D$:
\begin{gather}
\tilde{R}(S) \leq \sum_{i =1} ^{|S|-1} I\left(U_{\pi_i};
U_{\pi_{i+1}}^
{\pi_{|S|}} \right)  + I\left(U(S); U(S^c)\right) \nonumber \\
+ I\left(U(S); Z |U(S^c) \right), \forall S \subseteq \{1,  \cdots, L \},
 \label{general correlated MAC region}
\end{gather}

\no where $\tilde{R}(S) =
\sum_{i \in S} \tilde{R}_i$, $U(S) = \{U_i, i \in S \}$, $S^c$ is
the complementary set of $S$, $\{ \pi_1, \cdots, \pi_{|S|}\}$ is
an arbitrary permutation of $S$, and
 $U_{\pi_{i+1}}^{\pi_{|S|}} = \left[ U_{\pi_{i+1}}, \cdots,
U_{\pi_{|S|}}\right]$.
\end{theorem}

\begin{proof}
The proof follows the footsteps of the one proving achievability of the capacity region for
regular MAC channel with independent channel codewords.  The difference here is that we need to take
into account the correlations of the channel inputs from each user when the joint typical
sequence technique is used to compute the upper bound of the probability of various error events. The details
are deferred to the Appendix~\ref{appendix1}.
\end{proof}

It can be easily seen  that when inputs to the channel are
independent, the first and second terms in (\ref{general correlated MAC region}) vanish and consequently 
the inequality reduces to the one characterizing the capacity region for MAC channels with
independent inputs \cite[Chap.15.3]{cover_book}.

Next, we prove a sequence of lemmas in order to establish a connection between the JSCC and
SSCC approaches.
\begin{lemma} \label{lemma mutual0}

Given $U_j = X_j + V_j, j=1,2$ as in (\ref{theorem2:1}), $U_2 \rightarrow X_2 \rightarrow
X_1 \rightarrow U_1$ forms a Markov chain. As a result, we have

\begin{gather}
I(U_2;X_2 |U_1) = I(U_2;X_2)- I(U_1;U_2) \label{mutual I1} \\
I(U_1;X_1 |U_2) = I(U_1;X_1)- I(U_2;U_1) \label{mutual I2} \\
 I(U_1,U_2; X_1, X_2) = I(X_1;U_1)+I(X_2;U_2)-I(U_1;U_2)
 \label{mutual I3}
 \end{gather}
\end{lemma}
\begin{proof}
See Appendix~\ref{appendix2}.
\end{proof}

\begin{lemma} \label{lemma mutual1}
For  $U_j=X_j +V_j$, $j=1, \cdots, L$, the
following relation of mutual information holds
\begin{gather}
I\left(U(S); X(S) \right) =
\sum_{i\in S} I(U_i; X_i)- \sum_{i =1} ^{|S|-1} I\left(U_{\pi_i}; U_{\pi_{i+1}}^
{\pi_{|S|}} \right), \, \forall S \subseteq \{1, \cdots, L\}.
\end{gather}
\end{lemma}
\begin{proof}

WLOG, consider $S = \{1, 2, \cdots, s\}$. Define ${\bf
\tilde{U}_2} = \left[U_2, U_3, \cdots, U_s \right]$ and  ${\bf
\tilde{X}_2} = \left[X_2, X_3, \cdots, X_s \right]$. Apparently,
${\bf \tilde{U}_2} \rightarrow {\bf \tilde{X}_2} \rightarrow X_1
\rightarrow U_1$ forms a Markov chain. From (\ref{mutual I3}), we
immediately obtain: \beeq I\left[ U(S); X(S)\right] = I(X_1;U_1) +
I({\bf \tilde{U}_2}; {\bf \tilde{X}_2})- I(U_1;{\bf \tilde{U}_2})
\eeeq

\no Using same idea, it can be shown that \beeq 
I({\bf \tilde{U}_2}; {\bf \tilde{X}_2}) = I(U_2;X_2) + I(U_3
\cdots, U_s; X_3, \cdots, X_s) - I(U_2; U_3^{s}),  \eeeq

\no where $ I(U_3 \cdots, U_s; X_3, \cdots, X_s)$ can be
decomposed in a similar manner. Such decomposition can be
conducted iteratively until we reach \beeq I(X_{s-1},X_s; U_{s-1},
U_s) = I(X_{s-1}; U_{s-1}) + I(U_s;X_s) - I(U_s; U_{s-1}) \eeeq

Combining all iterations yields the desired result:
\begin{gather}
I\left(U(S); X(S) \right) = \sum_{i=1}^s I(U_i; X_i)- \sum_{i =1}
^{s-1} I\left(U_{i}; U_{i+1}^ {{s}} \right)
\end{gather}

\no As the whole derivation does not rely on the exact order of
$\{1, \cdots, s\}$, we thus complete the proof of Lemma~\ref{lemma
mutual1}.
\end{proof}

\begin{lemma} \label{lemma mutual3}
For the same $U(S)$ and $X(S)$ as in Lemma~\ref{lemma mutual0}, we have
\begin{gather}
I\left(U(S); X(S) \right) - I\left(U(S); U(S^c)\right) =
I\left[ U(S); X(S) |U(S^c)\right]  \nonumber \\
= I\left[U(S); X_0 | U(S^c)\right] + \sum_{i \in S} I\left[ U_i;
X_i | X_0 \right] \label{mutual I5}
\end{gather}
\end{lemma}

\begin{proof}
See appendix
\end{proof}

\begin{theorem}

 When each sensor performs
independent vector quantization and subsequently transmits the
 resulting   scaled quantization vector through a Gaussian MAC
channel, to reconstruct the Gaussian source at fusion center with
distortion no greater than $D$, the necessary and sufficient
condition is for any subset $S \subseteq \{1, 2, \cdots, L\}$, the
following inequality holds
\begin{gather}
I\left[U(S); X_0 | U(S^c)\right] + \sum_{i \in S} I\left[ U_i; X_i
| X_0 \right]  \leq I\left[U(S); Z | U(S^c) \right] \label{theorem-J}
\end{gather}

\no where
\begin{gather}
\mbox{LHS} =  -\frac{1}{2} \log \left[ \frac{D_E}{\sigma_S^2} +
\sum_{i \in S^c} \frac{D_E}{\sigma_{N_i}^2} \left(1 - 2^{-2 r_i}
\right) \right] + \sum_{i \in S} r_i  \label{LHS}
\end{gather}

\no and
\begin{gather}
\mbox{RHS} = \frac{1}{2} \log \left\{ 1 + \frac{1}{\sigma_W^2}
{\bf \sqrt{g}}(S)^T {\bf Q_{\Sigma_S}} {\bf \sqrt{g} }(S)
\right\}\label{RHS}
\end{gather}

\no with ${\bf \sqrt{g}}(S)^T = \left[ \sqrt{g_i}, i \in S
\right]$ and $ {\bf Q_{\Sigma_S}} = {\bf \Sigma_S} - {\bf
\Sigma_{S,S^c}} {\bf \Sigma_{S^c}}^{-1}{\bf \Sigma_{S^c,S}}$. 
The
auto- and cross-covariance matrices ${\bf \Sigma_S}$, ${\bf
\Sigma_{S^c}}$, ${\bf \Sigma_{S,S^c}}$  and ${\bf \Sigma_{S^c,S}}$
are defined as in (\ref{cov_matrix}).

\end{theorem}

\begin{proof}

To construct an estimate of $X_0$ at a fusion center with distortion no greater than
$D$  is equivalent to requiring  that the minimum compression rate $\tilde{R}_i$
satisfies $\tilde{R}_i =
I(X_i;U_i)$, as required by local vector quantization, and
that  $\tilde{R}_i, i=1, \cdots, L$ are in the region determined in Theorem~\ref{theorem correlated}.
Consequently, the conditions are translated to
 \begin{gather} \sum_{i \in
S} I(X_i; U_i) \leq  \sum_{i =1} ^{|S|-1} I\left(U_{\pi_i};
U_{\pi_{i+1}}^
{\pi_{|S|}} \right) \nonumber \\
 + I\left(U(S); U(S^c)\right) + I\left(U(S); Z |U(S^c) \right)
\end{gather}

\no From Lemma~\ref{lemma mutual1} and Lemma~\ref{lemma mutual3},
this condition is equivalent to
\begin{gather} \label{final1}
I\left[U(S); X_0 | U(S^c)\right] + \sum_{i \in S} I\left[ U_i; X_i
| X_0 \right] \leq I\left[ U(S); Z | U(S^c) \right].
\end{gather}

\no Define $r_i = I\left(X_i; U_i |X_0 \right)$. Then it is straightforward
to show that the LHS of
(\ref{final1}) is  equal to that
in (\ref{LHS}) by computing the mean squared error of estimating $X_0$ using $U(S)$ or
$[U(S), U(S^c)]$ \cite{oohama_IT05}, which is
\beeq
E \left[ |X_0|^2 |U(S) \right] =  \left[ \frac{1}{\sigma_S^2} +
\sum_{i \in S} \frac{1}{\sigma_{N_i}^2} \left(1 - 2^{-2 r_i}
\right) \right]^{-1}
\eeeq

Given $ I\left[ U(S); Z | U(S^c) \right] = H \left[Z |U(S^c)
\right] - H\left[ Z | U(S), U(S^c) \right]$ and ${\bf U}$ and $Z$
are Gaussian random vector/variables, it is sufficient to get the
conditional variance of $Z$ given the vector $U(S^c)$. This can be
boiled down to finding the conditional variance of $\sum_{i \in S}
\sqrt{g_i}\gamma_i U_i$ given $U(S^c)$ as $Z = \sum_{i=1}^L \sqrt{g_i} \gamma_i U_i
+ W$.

Based on Theorem~3 in \cite{anderson_multi}, we have
\begin{gather}
\mbox{Cov} \left[ U(S)| U(S^c) \right] = {\bf \Sigma_S} - {\bf
\Sigma_{S,S^c}} {\bf \Sigma_{S^c}}^{-1}{\bf \Sigma_{S^c,S}}
\end{gather}

\no Therefore,
\begin{gather}
\mbox{Var}\left(\sum_{i \in S} \sqrt{g_i}\gamma_i U_i \right) = {\bf
\sqrt{g}}(S)^T {\bf Q_{\Sigma_S}} {\bf \sqrt{g} }(S).
\end{gather}

\no The entropy can thus be  computed accordingly yielding
\begin{gather}
H \left[Z |U(S^c) \right] = \frac{1}{2} \log\left[ 2 \pi e \left(
\sigma_W^2 +   {\bf \sqrt{g}}(S)^T {\bf Q_{\Sigma_S}} {\bf
\sqrt{g} }(S) \right) \right] \nonumber \\
H \left[Z |U(S^c), U(S) \right] = \frac{1}{2} \log\left( 2 \pi e
\sigma_W^2\right)
\end{gather}
 \no which leads to (\ref{RHS}), and hence completes the proof.
\end{proof}

\subsection{Relationship With Separate Source-Channel Coding
Approach}
\label{relationship}

If we look closely at (\ref{theorem-J}) and (\ref{LHS}), we can easily see
that the LHS of the achievable rate region for the JSCC approach
actually characterizes the rate-distortion region for Gaussian sources with
conditionally independent (CI) condition \cite{oohama_IT05}.

Under the CI assumption, distributed source coding at sensors
includes two steps. The first step is the same as in JSCC, in which  an independent vector
quantization for Gaussian source at each sensor is conducted with
respect to the observed signal ${\bf X_j} = \{X_j[i], i =1,
\cdots, n\}$, which generates a vector ${\bf U_j^k} = \{U_j^k[i], i
=1, \cdots, n\}$, $k \in \{1, \cdots, 2^{n\tilde{R}_j}\}$. In the second step,  those  indexes of
$k_j$  are further compressed using
Slepian-Wolf's random binning approach \cite{oohama_IT05,
venod_ISIT_04}. Consequently,  there are $2^{n R_j}$ bins for  sensor $j$, which
contain all representation vectors ${\bf U_j^k}$ of measurements
${\bf X_j}$.  It was shown in \cite{oohama_IT05} that $R_j$ satisfy:
$\sum_{j \in S} R_j \geq I\left[U(S); X_0 | U(S^c)\right] + \sum_{i \in S} I\left[ U_i; X_i
| X_0 \right]$, for all $S \subseteq \{1, 2, \cdots, L\}$ in order to restore $X$ remotely
with distortion no greater than $D$.

For SSCC, to send indexes of bins correctly to the fusion center,
independent Gaussian codewords $\{Y_j[i] \sim \mathcal{N}(0,
P_j), i=1, \cdots, n\}$ for $j=1, \cdots, L$  for each bin
index are generated at $L$ sensors. To ensure indexes are
correctly decoded at the fusion center, the rate tuple $\{R_i, i =1, \cdots, L\}$
should also be contained in the capacity region of
Gaussian MAC channel with independent channel inputs under power
constraints $\{P_j, j=1, \cdots, L\}$. The region   is characterized by
$ \sum_{i \in S} R_i \leq
\frac{1}{2} \log \left[1 + \sum_{j \in S} \frac{P_j g_j}{\sigma_W^2}
\right]$, for all $S \subseteq \{1, 2, \cdots, L\}$.

The data processing at the fusion center consists of  three phases. In
the first phase, channel decoding is performed to recover the
indexes of bins containing $\{ {\bf U_j^k}, j=1, \cdots, L\}$. In the
second phase, joint typical sequences  $\{ {\bf U_j^k} \}$ are
obtained from $L$  bins whose indexes are restored. In
the last phase, $\{ {\bf U_j^k} \}$ are linearly combined
to estimate the source vector $\{X_0[i]\}$  under the minimum mean squared error
(MMSE) criterion.

Under SSCC, we can therefore obtain the sufficient and necessary condition for restoring
$X_0$ with MSE no greater than $D$:
\begin{gather}
I\left[U(S); X_0 | U(S^c)\right] + \sum_{i \in S} I\left[ U_i; X_i
| X_0 \right]  \nonumber \\ \leq
\frac{1}{2} \log \left[1 + \sum_{j \in S} \frac{P_j g_j}{\sigma_W^2}
\right], \forall S \subseteq \{1, 2, \cdots, L\}.
\end{gather}

In general,  we cannot say which approach, JSCC or SSCC, is better in terms
of the size of rate region.
This can be seen more clearly when we look at a
particular case for $L=2$.  When there are only two sensors,
to reconstruct  $\{X_0[i]\}$ with a distortion no greater than $D$
using JSCC or SSCC proposed as above,
the transmission powers $P_1$ and $P_2$, as well as $r_1$ and
$r_2$ satisfy:

\begin{gather}
r_1 -\frac{1}{2}
\log \left\{\frac{D_E}{\sigma_S^2} +
\frac{D_E}{\sigma_{N_2}^2}\left(1-2^{-2 r_2}\right) \right\} \nonumber \\ \leq
\frac{1}{2}
\log\left( 1 + \frac{P_1 g_1 ( 1- \tilde{\rho}_{1,2}^2)}{\sigma_W^2}\right)
\label{1}\\
r_2 -\frac{1}{2}
\log \left\{\frac{D_E}{\sigma_S^2} +
\frac{D_E}{\sigma_{N_1}^2}\left(1-2^{-2 r_1}\right) \right\} \nonumber \\ \leq
\frac{1}{2}
\log\left( 1 + \frac{P_2 g_2 ( 1- \tilde{\rho}_{1,2}^2)}{\sigma_W^2}\right) \label{2}
\\
 r_1 + r_2 + \frac{1}{2} \log \left(
\frac{\sigma_S^2}{D_E} \right)  \nonumber \\ \leq \frac{1}{2}
\log\left( 1 + \frac{P_2 g_2 + P_1 g_1 + 2 \tilde{\rho}_{1,2}
\sqrt{P_1 g_1 P_2 g_2}}{\sigma_W^2}\right) \label{3}
\end{gather}

\no where $\tilde{\rho}_{1,2}$ denotes the covariance coefficient between $U_i$ and $U_j$,
which is zero for SSCC and
\beeq \label{new rho 12}
\tilde{\rho}_{1,2} = \sqrt{\frac{q_1 q_2}{(1+q_1)(1+q_2)}}, \eeeq

\no for JSCC, as obtained in (\ref{new rho}) with
$r_i$ satisfying
\beeq  \label{two sensors r}
1/D_E = \frac{1}{\sigma_S^2} + \sum_{k=1}^{2}
\frac{1-2^{-2r_k}}{\sigma_{N_k}^2} \geq \frac{1}{D}.
\eeeq

It can be easily seen from (\ref{1})-(\ref{3}) that inequalities of
(\ref{1}) and (\ref{2}) under JSCC
are dominated by those under SSCC, i.e. $\{P_j, r_j\}$ satisfying (\ref{1})
and (\ref{2}) under
JSCC also satisfies the corresponding inequalities under SSCC,  while the
inequality (\ref{3}) under JSCC dominates that under SSCC.

To compare SSCC and JSCC, we next formulate a constrained  optimization problem in
which the objective is to minimize   the total transmission power of
$L$ sensors with a constraint that the distortion in restoring $X$ is no
greater than $D$.  For $L=2$, the problem can be stated as\\
\beeq  \label{opt problem}
\min_{P_i, r_i, i =1,2} P_1 + P_2, \mbox{ subject to (\ref{1})-(\ref{3}) and (\ref{two
sensors r}).}
\eeeq

\no which becomes power/rate allocations for SSCC and JSCC,
respectively, for different correlation coefficients $\tilde{\rho}$. The optimization results for SSCC and
JSCC under different channel and measurement parameters will reveal to us the relative efficiency of SSCC
and JSCC, which will be further compared with that for an uncoded scheme, as investigated in the next few
sections.


\section{Optimal Power and Rate Allocations to Minimize the Total
Transmission Power} \label{optimization}

\subsection{Geometric Programming Solution to Power/Rate Allocations}
\label{GP}

The constrained optimization problems in (\ref{opt problem}) are
non-convex. They can, however,  be solved efficiently  using standard techniques in convex
optimization by transforming the original problems into
relaxed convex geometric programming problems \cite{boyd_book}.

In this section, we take SSCC as an example to demonstrate how it works. For SSCC with
$\tilde{\rho}=0$, the rate tuple
$(R_1, R_2)$  should be taken from the
boundary of the capacity region for two-user Gaussian MAC channels to minimize $P_1+P_2$.
Consequently,
\begin{gather} R_1 = \frac{\alpha}{2}  \log\left(\frac{\sigma_W^2 + P_1
g_1}{\sigma_W^2} \right) + \frac{1-\alpha}{2} \log\left(
1+ \frac{P_1 g_1}{g_2 P_2 + \sigma_W^2} \right)  \nonumber \\
 R_2 = \frac{\alpha}{2}  \log\left( 1+
\frac{P_2 g_2}{g_1 P_1 + \sigma_W^2} \right) + \frac{1-\alpha}{2}
\log\left(\frac{\sigma_W^2 + P_2 g_2}{\sigma_W^2} \right) \label{mac
boundary}
\end{gather}

\no where $\alpha \in [0,1]$ is a time sharing factor.

Define $y_j = 2^{2R_j}$, $z_j = 2^{2 r_j}$ for $j=1,2$. We can
transform this  total power minimization problem for SSCC with $L=2$ to  an equivalent
generalized Signomial Programming problem \cite{boyd_tutorial}:
\begin{gather}
\min_{ P_j, y_j, z_j, j=1,2} P_1 + P_2, \, \,  \mbox{subject to:} \label{objective} \\
 \left(g_2 P_2 + \sigma_W^2\right)^{(1-\alpha)} y_1
\leq \nonumber \\
\left( 1 + \frac{g_1 P_1}{\sigma_W^2} \right)^{\alpha} \left(
g_1 P_1 + g_2
P_2 + \sigma_W^2 \right)^{(1-\alpha)}  \label{1,1} \\
 \left(g_1 P_1
+ \sigma_W^2\right)^{\alpha} y_2 \leq \nonumber \\
\left( 1 + \frac{g_2
P_2}{\sigma_W^2} \right)^{(1-\alpha)} \left( g_1 P_1 + g_2 P_2 +
\sigma_W^2 \right)^{\alpha}  \label{1,2} \\
\frac{\sigma_s^2}{\sigma_{N_2}^2 + \sigma_s^2} z_2^{-1} + D^{-1}
\frac{\sigma_s^2 \sigma_{N_2}^2}{\sigma_{N_2}^2 +
\sigma_s^2}y_1^{-1}z_1 \leq 1 \label{1,3} \\
\frac{\sigma_s^2}{\sigma_{N_1}^2 + \sigma_s^2} z_1^{-1} + D^{-1}
\frac{\sigma_s^2 \sigma_{N_1}^2}{\sigma_{N_1}^2 +
\sigma_s^2}y_2^{-1}z_2 \leq 1   \label{1,4} \\
y_1^{-1}y_2^{-1} z_1 z_2 \sigma_s^2/D \leq 1  \label{1,5} \\
D^{-1} + \sigma_{N_1}^{-2} z_1^{-1} + \sigma_{N_2}^{-2} z_2^{-1}
\leq \sigma_s^{-2} +  \sigma_{N_1}^{-2} +  \sigma_{N_2}^{-2}
\label{1,6}
\end{gather}

\no where constraints (\ref{1,1}) and (\ref{1,2}) are obtained by
relaxing equality constraints in (\ref{mac boundary}), and
constraints (\ref{1,3})-(\ref{1,6}) result from the transformation
of (\ref{1})-(\ref{3}), which 
are in the form of $f(x) \leq 1$, where $f(x)$ is a posynomial
function of $n$ variables \cite{boyd_book}:
$f(x) = \sum_{k=1}^K c_k x_1^{a_{1k}}
x_2^{a_{2k}} \cdots x_n^{a_{nk}}$,
where $c_k \geq 0$ and $x_j > 0$ for $j=1, \cdots, n$ and
$a_{ij} \in {\mathcal R}$.  In addition, constraints (\ref{1,1})
and (\ref{1,2}) are in the form of  generalized signomial
functions \cite{boyd_tutorial, chiang_tutorial} with fractional
powers.

Single condensation technique \cite{boyd_tutorial,
chiang_tutorial} can then be applied to convert this Signomial
programming problem to a standard geometric programming (GP)
problem. In this method, we replace $(1+P_1 g_1/\sigma_W^2)$ in
the RHS of (\ref{1,1}) by its geometric mean
$\beta_{11}^{\beta_{11}} \left(\frac{g_1 P_1}{\sigma_W^2
\beta_{12}}\right)^{\beta_{12}}$, and similarly $(1+P_2
g_2/\sigma_W^2)$ in the RHS of (\ref{1,2}) by
$\beta_{21}^{\beta_{21}} \left(\frac{g_2 P_2}{\sigma_W^2
\beta_{22}}\right)^{\beta_{22}}$, where $\beta_{i,j} \geq 0$ and
$\beta_{i,1}+\beta_{i,2} =1$ for $i=1,2$. In addition, we also
replace $(g_1 P_1 + g_2 P_2 + \sigma_W^2)$ by its geometric mean:
$\left(\frac{\sigma^2}{\gamma_1}\right)^{\gamma_1}
\left(\frac{g_1 P_1}{\gamma_2}\right)^{\gamma_2}\left(\frac{g_2
P_2}{\gamma_3}\right)^{\gamma_3}$,  where $\gamma_i \geq 0$ and $\sum_{i=1}^3
\gamma_i =1$.
Finally, to handle fractional powers in the LHS of (\ref{1,1}) and
(\ref{1,2}),  we introduce two auxiliary variables $t_1$ and $t_2$
to replace $g_1 P_1  + \sigma_W^2$ and $g_2 P_2 + \sigma_W^2$,
respectively, in the LHS of (\ref{1,1}) and (\ref{1,2}).
Accordingly, two additional posynomials are introduced on the list
of constraints: $g_i P_i + \sigma_W^2 \leq t_i$ for $i=1,2$.

The resulting standard geometric programming problem can thus be
solved in an iterative manner by repeatedly updating normalization
coefficients $\beta_{ij}$ and $\gamma_i$, and applying interior
point method for a given vector of these coefficients
\cite{boyd_tutorial, chiang_tutorial}.

Using the similar method, we can also transform the optimization
problem for the JSCC approach to a convex Geometric programming
problem. The details are skipped here \cite{wei_report_07}.

\subsection{Optimal Source/Channel Decoding Order for Non-Symmetric
Channels} \label{decoding order}

Although the optimization problems formulated in  (\ref{opt problem}) 
can only be solved algorithmically, we can still manage
to obtain some insights by scrutinizing the problem structures. In
this section, we will reveal some relationships between the
optimal decoding order and channel attenuation factors for
non-symmetric networks.

\subsubsection{Separate Source and Channel Coding} \label{separate
order}

We first show  the optimal source encoding/decoding order, as well
as channel decoder order for SSCC is uniquely determined by the
ordering of channel attenuation factors $\{g_i, i =1, \cdots,
L\}$, and has nothing to do with the ranking of sensor measurement
noise power $\{\sigma_{N_i}^2, i =1, \cdots, L \}$.

\begin{theorem}
\label{theorem sep order}
For SSCC, let $\pi^*$ denote any permutation of $\{1, \cdots, L \}$ such
that $g_{\pi^*(1)} \leq  g_{\pi^*(2)} \leq \cdots \leq g_{\pi^*(L)}$. To
minimize the total transmission power, the optimal decoding order for
channel codes at receiver is in the reversed order of $\pi^*$, i.e.
interference cancellation is in the order $\pi^*(L), \pi^*(L-1), \cdots,
\pi^*(1)$, which is also the decoding order of distributed source
codewords.
\end{theorem}

\begin{proof}
The proof consists of two steps. First, we will determine the channel
decoding order for a given vector of source encoding rates $\{R_i, i=1,
\cdots, L\}$. For SSCC, the rate tuple $\{R_i\}$ satisfies $\sum_{i \in S}
R_i \leq   \frac{1}{2} \log \left( 1 + \sum_{i \in S} P_i g_i /\sigma_W^2
\right) $, which is equivalent to
\beeq \label{contra modular}
\sum_{i\in S} X_i \geq f(S) \stackrel{\Delta}{=} \prod_{i \in S} 2^{2 R_i} -1, \forall
S \subseteq \{ 1, \cdots, L\}.
\eeeq

\no where $X_i = P_i g_i/\sigma_W^2$ and $f: 2^{E} \rightarrow \mathcal{R}_+ $ is a set
function with $E \stackrel{\Delta}{=} \{ 1, \cdots, L \}$.
Given $\{R_i\}$, the optimization problem then becomes $\min \sum_{i =1}^L \sigma_W^2
X_i/g_i$, subject to (\ref{contra modular}).

Based on the Corollary 3.13 in \cite{Tse-Hanly-I-98}, the set of power
vectors  $\{X_i\}$ satisfying (\ref{contra modular}) is a contra-polymatroid
$\mathcal{G}(f)$, as $f$
satisfies (1) $f(\phi) =0$ (2) $f(S) \leq f(T)$ if $S \subset T$ (3) $f(S) + f(T) \leq f(S
\cup T) + f(S \cap T) $.
From  Lemma 3.3 in \cite{Tse-Hanly-I-98}, the minimizing vector $\{X_i, i \in E\}$ for
$\min \sum_{i=1}^L \sigma_W^2 X_i/g_i$  is a vertex point $\{ X_{\pi^*(i)}\}$ of $\mathcal{G}(f)$, where $\pi^*$ is a permutation
on the set $E$ such that $1/g_{\pi^*(1)} \geq  \cdots \geq 1/g_{\pi^*(L)}$ and
\begin{gather}
X_{\pi^*(i)} =  f\left( \{ \pi^*(1), \cdots, \pi^*(i)\} \right)-
f\left( \{ \pi^*(1), \cdots, \pi^*(i-1)\} \right) \nonumber \\
=   \prod_{j=1}^i 2^{2 R_{\pi^*(j)}} -\prod_{j=1}^{i-1} 2^{2 R_{\pi^*(j)}}
\label{sep power vector}
\end{gather}

\no which thus proves the first part of this Theorem.

We next use (\ref{sep power vector}) to transform the objective function to
\beeq \label{new obj sep}
\sum_{i=1}^L \sigma_W^2 X_i/g_i  = \sum_{i=1}^{L-1} \left( \frac{1}{g_{\pi^*(i)}}-
\frac{1}{g_{\pi^*(i+1)}} \right)  \prod_{j=1}^i 2^{2 * R_{\pi^*(j)}}
\eeeq

For SSCC, rate tuples $\{ R_i, i =1, \cdots, L \}$ satisfy $\sum_{i \in S} R_i \geq I\left(
X(S); U(S) | U(S^c) \right)$. It is therefore quite straightforward to show that in order to
minimize the total transmission power in (\ref{new obj sep}), we need to have $\sum_{j=1}^i
R_{\pi^*(j)}$ achieve the lower bound, i.e.
\beeq  \sum_{j=1}^i R_{\pi^*(j)} = I\left(
X_{\pi^*(1)}^{\pi^*(i)};  U_{\pi^*(1)}^{\pi^*(i)}| U_{\pi^*(i+1)}^{\pi^*(L)} \right)
\eeeq

\no which implies that the decoding order for source codewords is  $\pi^*(L), \cdots,
\pi^*(1)$, independent of the ordering of variances $\{\sigma_{N_i}, i=1, \cdots, L\}$ of
measurement noise.

\end{proof}

Theorem~\ref{theorem sep order} implies  that sensor $\pi^*(L)$ does not conduct random
binning and its  quantization vector $\{U_{\pi^*(L)}[n]\}$ is
restored first.  $\{U_{\pi^*(L)}[n]\}$ is then used as side information to restore sensor
$\pi^*(L-1)$'s source codeword $\{U_{\pi^*(L-1)}[n]\}$ from this node's  first stage Gaussian source vector quantization,
which resides  in the bin whose index is decoded from the channel decoding step. This process continues until sensor
$\pi^*(1)$'s  first stage quantization vector $\{U_{\pi^*(1)}[n]\}$is restored by using all other
sensors' quantization as side information.

In the next section, we will see a similar conclusion can be
reached  for JSCC.

\subsubsection{Joint Source and Channel Coding} \label{joint order}

Unlike in the SSCC case where we have a nice geometric (contra-polymatroid) structure which enables us to
reach a conclusion valid for any arbitrary asymmetric networks, JSCC in general lacks such a feature for us
to exploit. We will instead , in this section,  focus on a case with only $L=2$
sensor nodes and establish a similar result as in
Section~\ref{separate order} for optimal channel decoding orders.
WLOG, we assume $g_1>g_2$ in the
subsequent analysis. 

\begin{theorem} \label{theorem4}
Given a pair of quantization rates $\tilde{R}_1$ and
$\tilde{R}_2$, when $g_1>g_2$, the optimal decoding order to
minimize the total transmission power $P_1 + P_2$ is to decoder
node $1$'s signal first, and then node $2$'s information after
removing the decoded node $1$'s signal from the received signal,
i.e.
\begin{gather}
\tilde{R}_1 = I(U_1; Z ) = \frac{1}{2} \log \left( 1 +
\frac{\left(\sqrt{P_1 g_1 } + \tilde{\rho} \sqrt{P_2 g_2}
\right)^2}{\sigma_W^2 + P_2 g_2 ( 1- \tilde{\rho}^2)}\right) \nonumber \\
\tilde{R}_2= I(U_2;Z, U_1)
 = \frac{1}{2} \log
\left( \frac{P_1 (1 - \tilde{\rho}^2)+\sigma_W^2}{\sigma_W^2 ( 1-
\tilde{\rho}^2)}\right) \label{eq: theorem4}
\end{gather}

\no where $\tilde{\rho}$ is the same as in \ref{new rho 12}.
\end{theorem}
\begin{proof}
See appendix~\ref{JSCC order}.
\end{proof}

\section{Uncoded Sensor Transmission in Fusion} \label{uncoded}

For  Gaussian sensor networks as modeled in Section~\ref{model},
it has been shown recently \cite{gastpar_IT05, gastpar_JSAC_05}
that uncoded transmission, i.e. each sensor only forwards a scaled
version of its measurements to the fusion center, asymptotically
achieves the lower bound on distortion when the number of sensors
grow to infinity and system is symmetric. In the context of the
theme of this paper, we, in this section, investigate the optimal
power allocation strategy when a finite number of sensors deploy
the uncoded scheme under  more general channel conditions.

For uncoded transmission, the transmitted signal by node $j$ is
$Y_j[i] = \alpha_j X_j[i]$, where $\alpha_j =
\sqrt{\frac{P_j}{\sigma_S^2 + \sigma_{N_j}^2}}$ is a scaling
factor to make the transmission power $E|Y_j[i]|^2 = P_j$. The
received signal at the fusion center is therefore $Z[i] =
\sum_{j=1}^L Y_j[i] \sqrt{g_j} + W[i]$. The linear MMSE estimate
of $X_0[i]$ is: $\hat{X}_0[i] = \gamma Z[i]$, where the
coefficient $\gamma$ can be obtained using Orthogonal principle:
$E\left[\left(X_0[i]- \hat{X}_0[i]\right) Z[i]\right] =0$.
The resultant MSE is \beeq \label{mse uncoded}
E\left|X_0[i]- \hat{X}_0[i]\right|^2 = \sigma_S^2
\frac{\sum_{j=1}^L \frac{P_j g_j \sigma_{N_j}^2}{\sigma_S^2 +
\sigma_{N_j}^2}+ \sigma_W^2}{\sum_{j=1}^L P_j g_j + B +
\sigma_W^2} \eeeq

\no where  $ B = \sum_{i=1}^L \sum_{i\neq j, j =1 }^L \rho_{i,j}
\sqrt{P_i g_i} \sqrt{P_j g_j}$ and the covariance coefficients
$\rho_{i,j}$ is the same as in (\ref{direct rho}). 

When $L=2$, the power control problem under a distortion
constraint $E\left|X_0[i]- \hat{X}_0[i]\right|^2  \leq D$ for the
uncoded scheme can be formulated as:
\begin{gather}
\min P_1 + P_2\;\;\mbox{subject to:} \nonumber \\ \frac{1}{2} \log
\left[ \frac{\sigma_S^2}{D} \left( 1 +
\frac{\sigma_{N_1}^2}{\sigma_S^2+ \sigma_{N_1}^2} \frac{P_1
g_1}{\sigma_W^2} + \frac{\sigma_{N_2}^2}{\sigma_S^2+
\sigma_{N_2}^2} \frac{P_2
g_2}{\sigma_W^2}\right) \right] \nonumber \\
\leq \frac{1}{2} \log \left[1 + \frac{P_1 g_1 +P_2 g_2 + 2
\rho_{1,2} \sqrt{P_1 g_1 P_2 g_2}}{\sigma_W^2} \right]
\label{uncoded opt}
\end{gather}

\no This problem can again be transformed to a GP problem using
the condensation technique applied in Section~\ref{GP}. We
skip the details here.

What deserves our attention  is that when
we compare the constraint in (\ref{uncoded opt}) with
(\ref{3}),  there is a striking
similarity when we substitute $r_i$ with $\tilde{R}_i$, whose
relationship was introduced in (\ref{R with r}). After
substitution, (\ref{3}) becomes
\begin{gather}  \frac{1}{2} \log \left\{
\frac{\sigma_S^2}{D} \left[1 + A_1 \right] \left[1 +
A_2 \right] \right\} \nonumber \\
\leq \frac{1}{2} \log \left[1 + \frac{P_1 g_1 +P_2 g_2 + 2
\tilde{\rho}_{1,2} \sqrt{P_1 g_1 P_2 g_2}}{\sigma_W^2} \right]
\end{gather}

\no where $A_i = \frac{\sigma_{N_i}^2}{\sigma_S^2+
\sigma_{N_i}^2} ( 2^{2 \tilde{R}_i}-1)$ for $i=1,2$, and  $\tilde{\rho}_{1,2}$
and $\rho_{1,2}$ are associated as
in (\ref{two rhos}).

For JSCC,  we have $\tilde{R}_i \leq I(U_i;Y
|U_j) + I(U_i;U_j)$, which is equivalent to \beeq 2^{\tilde{R}_i}-
\frac{1}{1 - \tilde{\rho}_{1,2}^2} \leq \frac{P_i g_i}{\sigma_W^2}
\eeeq

\no We can infer from (\ref{1})-(\ref{3}), as well as
(\ref{uncoded opt}) that it is in general hard to argue which
approach is the most energy efficient in terms of the total power
consumption under a common distortion constraint $D$,  which will
be further exemplified in our simulation results in
Section~\ref{simulation}. There, we will see the most energy
efficient approach depends on exact values of $\sigma_{N_j}^2$, as
well as $g_i$ and $D$.

However, when a system becomes homogeneous and symmetric in the sense that
$\sigma_{N_i}^2 = \sigma_{N_j}^2$ and $g_i = g_j$ for all $i,j \in
\{ 1, \cdots, L\}$, we have consistent results for both finite
number of $L$ and   asymptotically large $L$, as revealed in
the next section, when we compare these three approaches.

\section{Energy Consumption Comparison for Homogeneous Networks: Finite and
Asymptotic Results} \label{asymptotic comparision}

In this section, we provide analytical results on comparisons
between different transmission strategies proposed thus far, including JSCC, SSCC and
uncoded schemes in terms of
their total transmission power consumptions when system becomes
homogeneous.

\subsection{Comparison under finite $L$ and finite SNR}

\begin{theorem} \label{theorem5}
When a system of $L< \infty$ sensors  becomes symmetric, the total
power consumption for the separate, joint and uncoded schemes
proposed previously follow: \beeq \label{relative order}  \left( \frac{P
g}{\sigma_W^2}\right)_{LoB}< \left(\frac{P g}{\sigma_W^2}\right)_A
< \left(\frac{P g}{\sigma_W^2}\right)_J < \left(\frac{P
g}{\sigma_W^2}\right)_S \eeeq

\no where $\left( \frac{P g}{\sigma_W^2}\right)_{LoB}$ is the
lower-bound on transmission power, and $g$ is the common channel
gain from each sensor to the fusion center. Indexes A, J and S
represent the uncoded, joint and separate encoding  schemes,
respectively.
\end{theorem}

\begin{proof}

The proof hinges upon the analysis of rate and power allocations
for symmetric networks for  different schemes.

\vspace{0.1in} \no {\em \bf Separate Coding:}

We first look at the separate source and channel coding approach.
In symmetric networks, for source coding part, each node employs
identical compression rate $R$, which satisfies $LR = Lr +
\frac{1}{2}\log\frac{\sigma_S^2}{D}$, where $r$ is the solution to
\beeq \label{symmetric r} \frac{1}{\sigma_S^2} +
\frac{L}{\sigma_{N_1}^2}( 1- 2^{-2 r}) = \frac{1}{D}. \eeeq

\no As a result \cite{oohama_IT05}, \beeq \label{separate
symmetric} LR = -\frac{L}{2} \log \left( 1 -
\frac{\sigma_{N_1}^2}{L}\left( \frac{1}{D}- \frac{1}{\sigma_S^2}
\right) \right) + \frac{1}{2} \log\frac{\sigma_S^2}{D} \eeeq

\no  For channel coding part, to minimize the total transmission
power, it is optimal to let each sensor transmit at the same power
$P$ and same  rate $R$ which can be  achieved by jointly decoding
all node's information at the fusion center. Therefore, the total
compression rate also satisfies \beeq \label{separate 2} LR =
\frac{1}{2} \log \left( 1 + \frac{LP g}{\sigma_W^2}  \right)\eeeq

\no Combining (\ref{separate symmetric}) and (\ref{separate 2})
yields
\be \label{separate 3} \left( \frac{P
g}{\sigma_W^2}\right)_S & = & -\frac{1}{L} +
\frac{\sigma_S^2}{LD}\left( 2^{2r}\right)^L \nonumber \\ & = &
-\frac{1}{L} + \frac{\sigma_S^2}{LD} \left[ 1- \left(
\frac{1}{D}-\frac{1}{\sigma_S^2}\right) \frac{\sigma_{N_1}^2}{L}
\right]^{-L} \ee

\no {\em \bf Joint Coding:}

For the joint source-channel coding scheme, the vector
quantization rate for each sensor is equal to  $\tilde{R} =
I(X_1;U_1)$, which is associated with  $r$ as shown by
(\ref{theorem2:0}), where $r$ can be further obtained using
(\ref{symmetric r}).

By applying the techniques used in driving the constraints in
(\ref{1})-(\ref{3}) to the $L>2$  case when channel is symmetric,
the quantization rate and the resultant L-user multiple access
channel with $L$ correlated inputs $\{U_1, \cdots, U_L\}$ are
associated by  \be \label{joint entropy1} Lr +
\frac{1}{2} \log
\frac{\sigma_S^2}{D} & = & I(U_1, \cdots, U_L; Z) \nonumber \\
& = & H(Z) - H(Z|U_1, \cdots, U_L),
 \ee

\no where the conditional entropy $H(Z|U_1, \cdots, U_L) =
\frac{1}{2} \log\left( 2 \pi e \sigma_W^2 \right)$, and  the
entropy of $Z= \sum_{j=1}^L \sqrt{g} \gamma U_j + W$ is $H(Z)=
\frac{1}{2} \log \left[ 2 \pi e \left( \sigma_W^2 + g
\mbox{Var}(\gamma \sum_{j} U_j) \right)\right]$. Given the covariance
coefficient between $U_i$ and $U_j$: $\tilde{\rho} =
\frac{\sigma_S^2}{\sigma_S^2 + \sigma_{N_1}^2} ( 1- 2^{-2
\tilde{R}})$, we have \beeq  \label{joint entropy2} E|\gamma \sum_{j=1}^L
U_j |^2 = LP + (L^2-L) \tilde{\rho}P \eeeq

\no From (\ref{symmetric r}) and (\ref{theorem2:0}), we obtain
\beeq \label{joint entropy3} \tilde{\rho} = \left(
\frac{\sigma_S^2}{D}-1\right) \left( L + \frac{\sigma_S^2}{D}-1
\right)^{-1}.
\eeeq

The transmission power $P$ for joint coded scheme in symmetric
networks can thus be computed using (\ref{joint entropy1}) and
(\ref{joint entropy2}): \begin{gather}
\left(\frac{Pg}{\sigma_W^2}\right)_J  = \left[-\frac{1}{L} +
\frac{\sigma_S^2}{LD}\left( 2^{2r}\right)^L\right] \left( 1 +
\left(L- \frac{1}{L}\right) \tilde{\rho}\right)^{-1}\label{joint
entropy4}
\end{gather}

\no Comparing (\ref{joint entropy4}) with (\ref{separate 3}), it
is apparent that $ \left(\frac{Pg}{\sigma_W^2}\right)_J <
\left(\frac{Pg}{\sigma_W^2}\right)_S$.

\vspace{0.1in} \no {\em \bf Uncoded Scheme}

When sensor transmits scaled measurements in a symmetric network,
we can obtain the minimum transmission power $P$ by making the
mean squared error obtained in (\ref{mse uncoded}) equal to $D$
and substituting $P_j$, $g_j$ and $\sigma_{N_j}^2$ by $P$, $g$ and
$\sigma_{N_1}^2$, respectively. As a result, we obtain
\begin{gather}
\hspace{-0.2in} \left(\frac{Pg}{\sigma_W^2}\right)_A =  \left[ \frac{L^2
\sigma_S^2}{\sigma_S^2 + \sigma_{N_1}^2} - L \left(
\frac{\sigma_S^2}{D} -1 \right) \frac{\sigma_{N_1}^2}{\sigma_S^2 +
\sigma_{N_1}^2} \right]^{-1}  \left( \frac{\sigma_S^2}{D} -1
\right) \label{uncoded power}\end{gather}

To compare $\left(\frac{Pg}{\sigma_W^2}\right)_A$ with
$\left(\frac{Pg}{\sigma_W^2}\right)_J$, we need to introduce an
auxiliary variable. Define \beeq \label{inequality QL} \tilde{Q}_L
= 1 - \frac{\sigma_S^2 + \sigma_{N_1}^2}{\sigma_S^2} \tilde{\rho}
< 1 - \tilde{\rho} . \eeeq

\no We can therefore re-derive the minimum power for uncoded
scheme: \beeq \label{uncoded power2}
\left(\frac{Pg}{\sigma_W^2}\right)_A = \frac{1}{L} \left(
\frac{1}{\tilde{Q}_L} -1\right). \eeeq

In addition, we can express $2^{2r} = \left[ 1- \left(
\frac{1}{D}-\frac{1}{\sigma_S^2}\right) \frac{\sigma_{N_1}^2}{L}
\right]$ as \beeq 2^{2r} = \left[\frac{\tilde{Q}_L}{L} \left( L +
\frac{\sigma_S^2}{D}-1\right)\right]^{-1} = \frac{1
-\tilde{\rho}}{\tilde{Q}_L} >1 \eeeq

\no which is used to transform
$\left(\frac{Pg}{\sigma_W^2}\right)_J$ as \beeq \label{jointF}
\left(\frac{Pg}{\sigma_W^2}\right)_J = \frac{1}{L} \left\{ \left(
\frac{1 -
\tilde{\rho}}{\tilde{Q}_L}\right)^{L-1}\frac{1}{\tilde{Q}_L} -
\frac{L -1 + \sigma_S^2/D}{L \sigma_S^2/D} \right\}\eeeq

\no where the second term $\frac{L -1 + \sigma_S^2/D}{L
\sigma_S^2/D}<1$ due to $L>1$ and $\sigma_S^2 >D$. Since $1 -
\tilde{\rho} > \tilde{Q}_L$, after comparing (\ref{jointF}) and
(\ref{uncoded power2}), it is straightforward to show
$\left(\frac{Pg}{\sigma_W^2}\right)_J>
\left(\frac{Pg}{\sigma_W^2}\right)_A$.

\vspace{0.1in} \no {\bf Lower Bound of Transmission Power}

Since $X_0 \rightarrow \{ X_1, \cdots, X_L \} \rightarrow Z
\rightarrow \hat{X}_0$ forms a Markov chain, by Data Processing
Inequality \cite{cover_book}, we have $I(X_0; \hat{X}_0) \leq
I(X_1, \cdots, X_L; Z)$. On one hand, to ensure  $E|X_0 -
\hat{X}_0|^2 \leq D$, it can be shown  $I(X_0; \hat{X}_0) \geq
\frac{1}{2} \log \frac{\sigma_S^2}{D}$ using rate distortion
results \cite{cover_book}. On the other hand, the mutual
information $I(X_1, \cdots, X_L; Z)$ is upper-bounded by the
mutual information of an additive noise Gaussian channel with
channel gain $\sqrt{g}$ and total transmission power upper-bounded
by $ E |\sum_{j=1}^L X_j|^2 = L P + (L^2-L) \rho P $, where $\rho =
\frac{\sigma_S^2}{\sigma_S^2 + \sigma_{N_1}^2}$
is the covariance coefficient between $X_i$ and $X_j$ for $i \neq
j$.  Consequently, \beeq I(X_1, \cdots, X_L; Z) \leq \frac{1}{2}
\log \left[ 1 + \frac{Pg}{\sigma_W^2} \left(  L  + (L^2-L) \rho
\right) \right] \eeeq

From $ \frac{1}{2} \log \frac{\sigma_S^2}{D} \leq I(X_0;
\hat{X}_0) \leq I(X_1, \cdots, X_L; Z)$, we obtain the lower-bound
of transmission power: \beeq \label{lowerbound} \left( \frac{P
g}{\sigma_W^2}\right)_{LoB} =
\frac{\frac{\sigma_S^2}{D}-1}{L+(L^2-L) \rho} \eeeq

Compare this lower bound with (\ref{uncoded power}), we have
$\left( \frac{P g}{\sigma_W^2}\right)_{LoB} < \left( \frac{P
g}{\sigma_W^2}\right)_A$.

Therefore, we have shown that the order in (\ref{relative order}) holds and
thus completed  the proof.

\end{proof}

\subsection{Comparison under  $L \rightarrow \infty$ and finite SNR}

In this section, we provide the scaling behaviors of the total
transmission power of various schemes studied so far. In
particular, we are interested in how $\frac{P g}{\sigma_W^2}$
scales with respect to the number of sensors $L$ in a symmetric
system under a common constraint on distortion no greater than
$D$. The analysis is quite straightforward based on the results
for a finite $L$ number of sensors that we have obtained in
(\ref{separate 3}), (\ref{jointF}), (\ref{uncoded power2}) and
(\ref{lowerbound}), for separate, joint, uncoded schemes and the
lower bound, respectively.

\begin{theorem}
\begin{gather}
\lim_{L\rightarrow \infty} \left(\frac{L g P}{\sigma_W^2}\right)_S
= \frac{\sigma^2_S}{D} \exp \left[ \sigma_{N_1}^2\left(
\frac{1}{D} - \frac{1}{\sigma_S^2}\right) \right]-1
\nonumber \\
\lim_{L \rightarrow \infty} \left(\frac{L g
P}{\sigma_W^2}\right)_J = \exp \left[ \sigma_{N_1}^2\left(
\frac{1}{D} - \frac{1}{\sigma_S^2}\right) \right]
-\frac{D}{\sigma_S^2}
\nonumber \\
\lim_{L \rightarrow \infty} \left(\frac{L^2 g P}{\sigma_W^2}
\right)_A =\lim_{L \rightarrow \infty} \left(\frac{L^2 g
P}{\sigma_W^2} \right)_{LoB} \nonumber \\ = \left( \frac{1}{D} -
\frac{1}{\sigma_S^2}\right) \left(\sigma_S^2
+\sigma_{N_1}^2\right)
\end{gather}
\end{theorem}
\begin{proof}
The proof of these convergence results is quite straightforward based upon the results for
finite $L$ as above, and is skipped here.

\end{proof}

\no It is obvious that the transmission power of the uncoded
scheme shares the same scaling factor as the lower-bound, which is
in the order of $1/L^2$. The asymptotic optimality of the uncoded
scheme in symmetric Gaussian sensor networks is not a new result,
which has been attained previously in \cite{gastpar_IT05}
\cite{gastpar_IPSN03}. In \cite{gastpar_IT05, gastpar_IPSN03}, the
authors  assumed a fixed transmission power and showed that the
distortion achieved using uncoded approach has the same asymptotic
scaling law as that obtained via a lower bound. Here, we provide a
different perspective in assessing its optimality for the uncoded
scheme in terms of the total transmission power while meeting a
fixed distortion constraint.

Both joint and separate coding schemes have the scaling factor in
the order of $1/L$. Asymptotically, joint coding scheme saves in
total transmission power by a factor of $\sigma_S^2/D$ as compared
with the separated approach.

\subsection{Comparison under finite $L$ and  $\mbox{SNR} \rightarrow
\infty$}

Given the number of sensors $L<\infty$, we are interested in the scaling
factors associated with transmission SNR $P/\sigma_W^2$, mean squared error $D$ and
measurement noise variance $\sigma_N^2$ in homogeneous networks. In
particular, we need to investigate  how the following
asymptotic factors are related,  $\sigma_N^2
\rightarrow 0$, $D \rightarrow 0$ and $P/\sigma_W^2 \rightarrow \infty$.

We first need to identify the limit imposed upon the scaling factor related
with $\sigma_N^2$ and $D$. As can be seen from
both (\ref{symmetric r}) and (\ref{uncoded power}) under  SSCC,
JSCC and uncoded approaches, it is required that
\beeq \label{ineq 1}
\lambda\left(\sigma_N^2\right) \stackrel{\Delta}{=}
 \frac{\sigma_N^2}{L} \left(\frac{1}{D}-\frac{1}{\sigma_S^2} \right) \leq 1.
\eeeq

\no Denote $\gamma^* =
\lim_{\sigma_N^2 \rightarrow 0} \frac{\ln D}{\ln \sigma_N^2} $. It can be
seen that $\gamma^*$ has to satisfy $ \gamma^* \in [0, 1]$ in order to have
 the inequality in (\ref{ineq 1}) hold. Consequently,
\beeq \label{lambda*}
\lambda^* = \lim_{\sigma_N^2 \rightarrow 0} \lambda\left(\sigma_N^2\right)= \left\{
\begin{array}{ll}
0, & 0 \leq  \gamma^* < 1 \\
\frac{1}{L} & \gamma^* = 1
\end{array}
\right.
\eeeq

\begin{theorem} \label{asymptotic ratio}

The asymptotic ratios associated with $D$, $\sigma_N^2$ and $P/\sigma_W^2$
have the following relationship:
\begin{gather}
\lim_{\sigma_N^2 \rightarrow 0} \frac{\left(Pg/\sigma_W^2\right)_S}{\sigma_S^2/D} = \frac{1}{L
(1-\lambda^*)^L} \label{S} \\
\lim_{\sigma_N^2 \rightarrow 0}
\frac{\left(Pg/\sigma_W^2\right)_J}{\sigma_S^2/D} = \frac{1}{L^2
(1-\lambda^*)^L} \label{J} \\
\lim_{\sigma_N^2 \rightarrow 0}
\frac{\left(Pg/\sigma_W^2\right)_A}{\sigma_S^2/D} = \frac{1}{L^2
(1-\lambda^*)} \label{U}
\end{gather}
\end{theorem}

\begin{proof}
The proof follows straightforwardly with
(\ref{separate 3}),  (\ref{joint entropy4}) and  (\ref{uncoded power}) for
SSCC, JSCC and uncoded schemes, respectively, as we let $\sigma_N^2
\rightarrow 0$ and $\frac{\sigma_N^2}{D}  \approx L \lambda^*$.
\end{proof}

We can see from (\ref{S})-(\ref{U}) that when $\gamma^* \in [0, 1)$, i.e.
$\lambda^*=0$, the JSCC and uncoded schemes have the same aysmptotic
ratio between the received SNR and S-MSE-Ratio, which is smaller than that for  the
SSCC approach by a factor of  $1/L$.  If $\gamma^* =1$, i.e. $\lambda^*=1/L$,
uncoded approach has the smallest ratio among all schemes. However, if we
introduce the SNR exponent as defined in \cite{caire-SNR-exponent},
\beeq \label{SNR exp}
\eta \stackrel{\Delta}{=} - \lim_{\sigma_N^2 \rightarrow 0}
\frac{\ln D}{\ln \left(Pg/\sigma_W^2\right)},
\eeeq

\no All three approaches share the same ratio $\eta = 1$ for finite $L$.

Theorem~\ref{asymptotic ratio} therefore provides us another perspective to
compare these remote estimation approaches.     It demonstrates the proposed joint
source and channel coding scheme has potentially the same asymptotic
performance as the uncoded one in high channel SNR and high measurement SNR
regions for all ratio exponents $\gamma^* \in [0, 1)$.
In addition, speaking of SNR exponent of distortion measure in
the high SNR region, all three schemes investigated in this paper share the
same asymptotic ratio $\eta=1$.

An additional remark we next make is about  the limitation as to adopting SNR exponent $\eta$ as  a
metric to characterize the asymptotic performance in large SNR regions \cite{caire-SNR-exponent}.  It can be seen clearly
from the above analysis that the SNR exponent obscures the asymptotic difference between SSCC and
JSCC, as well as the uncoded approaches, which have different linear
ratios as shown in Theorem~\ref{asymptotic ratio}. These differences are gone once log-scale is
imposed, however.

Note also that  
we have implicitly  assumed that 
the spectral efficiency of the system model in this paper, which is the ratio
the source bandwidth over channel bandwidth \cite{caire-SNR-exponent},  is one. 


\section{Numerical Results} \label{simulation}
\begin{figure}[ht]
  \centering
         \centerline{
        \scalebox{0.8}
          {
        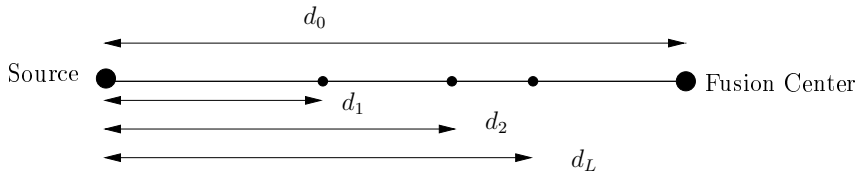
      }
 }
    \caption{1-D location Model}
 \label{topology}
\end{figure}

In this section, the three approaches proposed in this paper are examined and compared
with each other by looking at each of their optimal total transmission powers under the
constraint of restoring $X_0$ with MSE no greater than $D$, some prescribed threshold.
Particularly, we consider a linear network
topology  where a source, fusion center and $L=2$ sensor nodes are
located on a same line as illustrated in Figure~\ref{topology}.
To
associate positions of sensor nodes with channel gains and
measurement noise, we assume a path-loss model with coefficient
$\beta_s$ and $\beta_c$ for $g_i$ and $\sigma_{N_i}^2$,
respectively:  $g_i \propto 1/(d_0- d_i)^{\beta_c}$ and
$\sigma_{N_i}^2 \propto d_j^{\beta_s}$, for $i=1,\cdots, L$, where
$d_0$ is the distance between source and fusion center, and $d_i$
is the distance between the $i$-th sensor and source.
Given a distortion upper-bound $D< \sigma_S^2$ and $\beta_c=\beta_s = \beta$,
the distance between the source and fusion center has to satisfy the following
inequality, $d_0 <  (L)^{1/\beta} \left[\frac{1}{D} -
\frac{1}{\sigma_S^2}\right]^{-1/\beta}$, which is obtained by making the MSE
using $\{X_i\}$ to estimate $X_0$ no greater than $D$.

\begin{figure}[htb]
\vfill
\begin{minipage}[b]{1.0\linewidth}
  \centering
   \centerline{\epsfig{figure=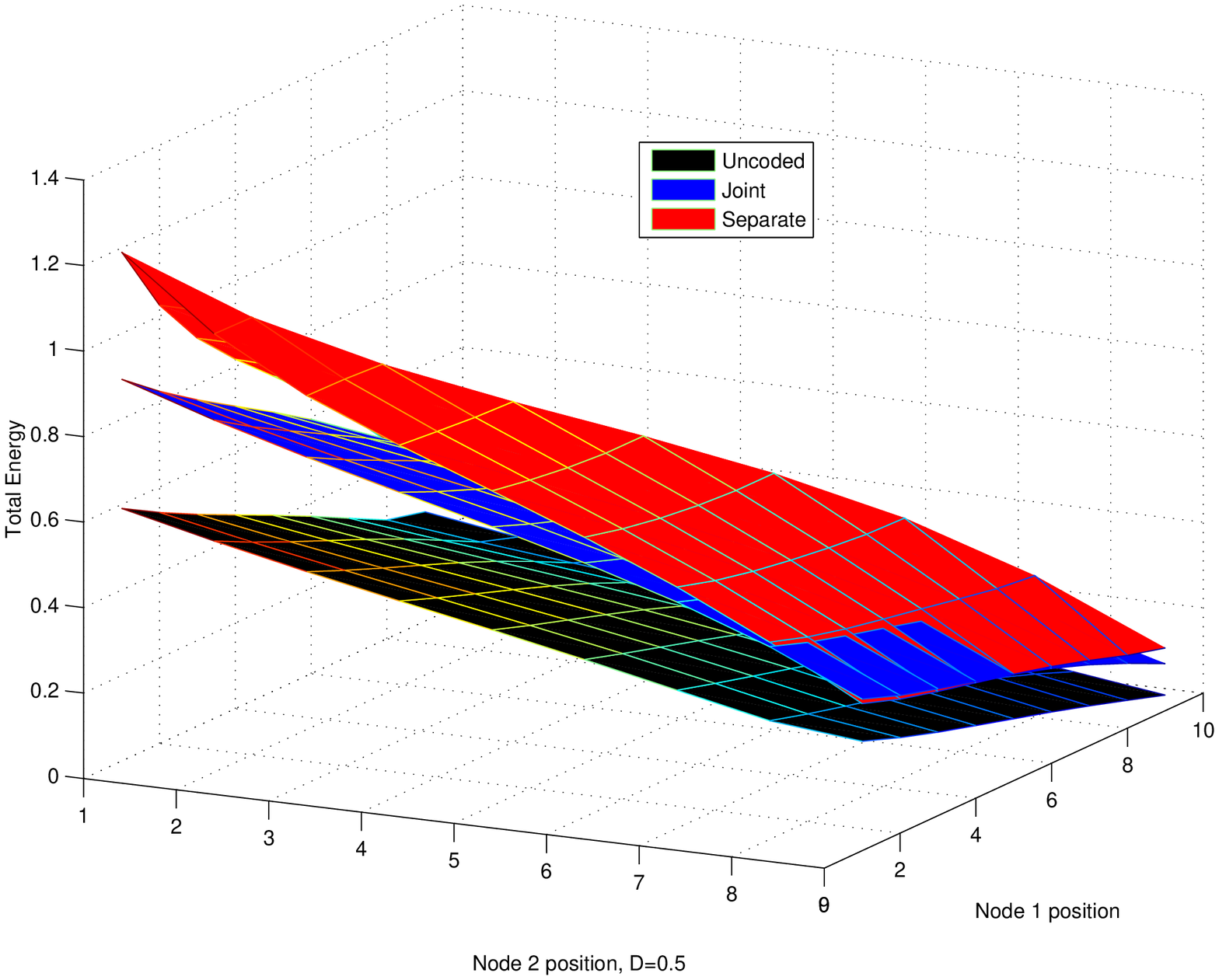,width=9.5cm}} 
  \centerline{(a) Result 1:  $D=0.5$, $\sigma_S^2 = \sigma_W^2=1$,
$\beta_c = \beta_s=2$.}\medskip
\end{minipage}
\begin{minipage}[b]{1.0\linewidth}
  \centering
\centerline{\epsfig{figure=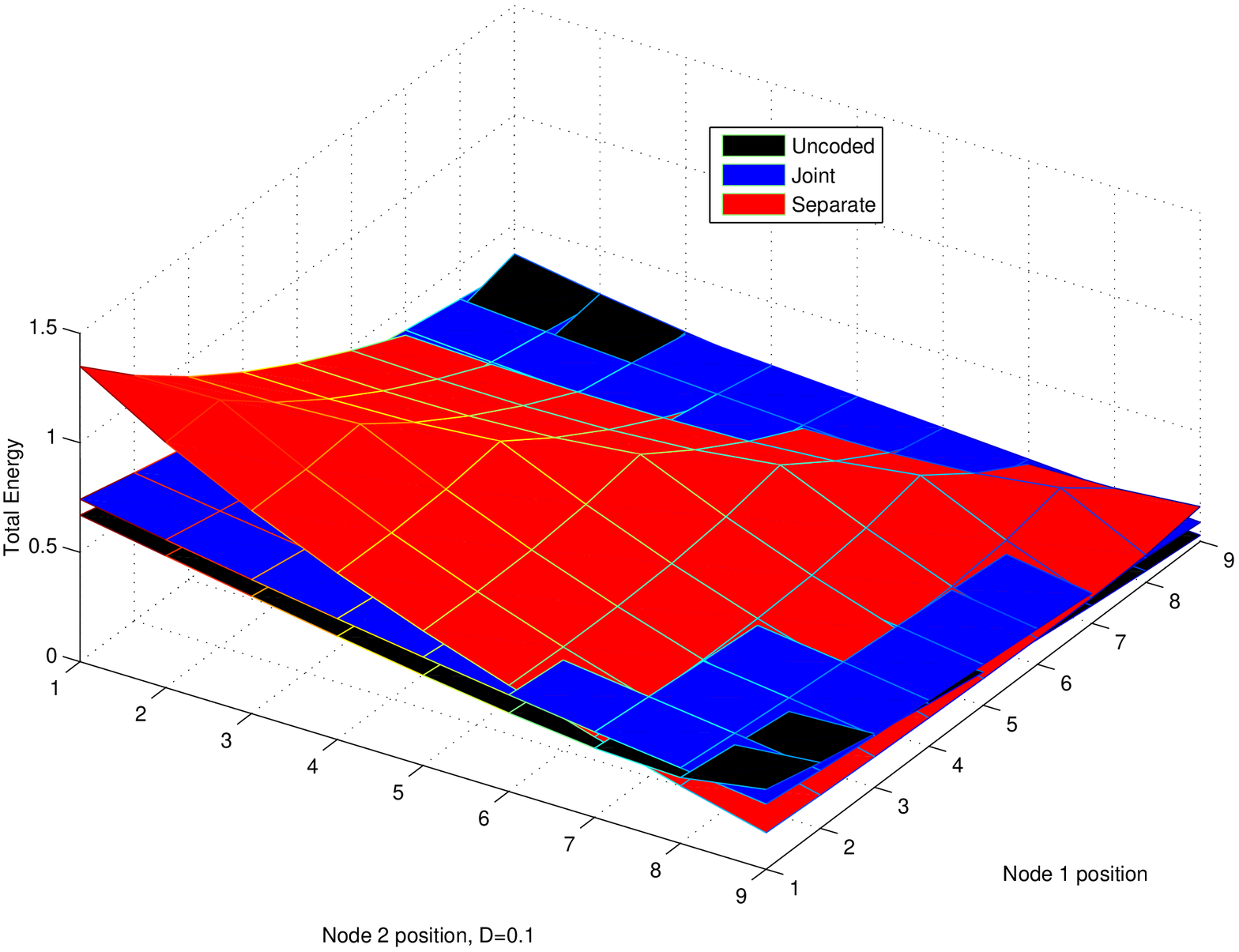,width=9.5cm}}
  \centerline{(b) Results 2: $D=0.1$, $\sigma_S^2 = \sigma_W^2=1$, $\beta_c
= \beta_s =2$}. \medskip
\end{minipage}
\begin{minipage}[b]{1.0\linewidth}
  \centering
\centerline{\epsfig{figure=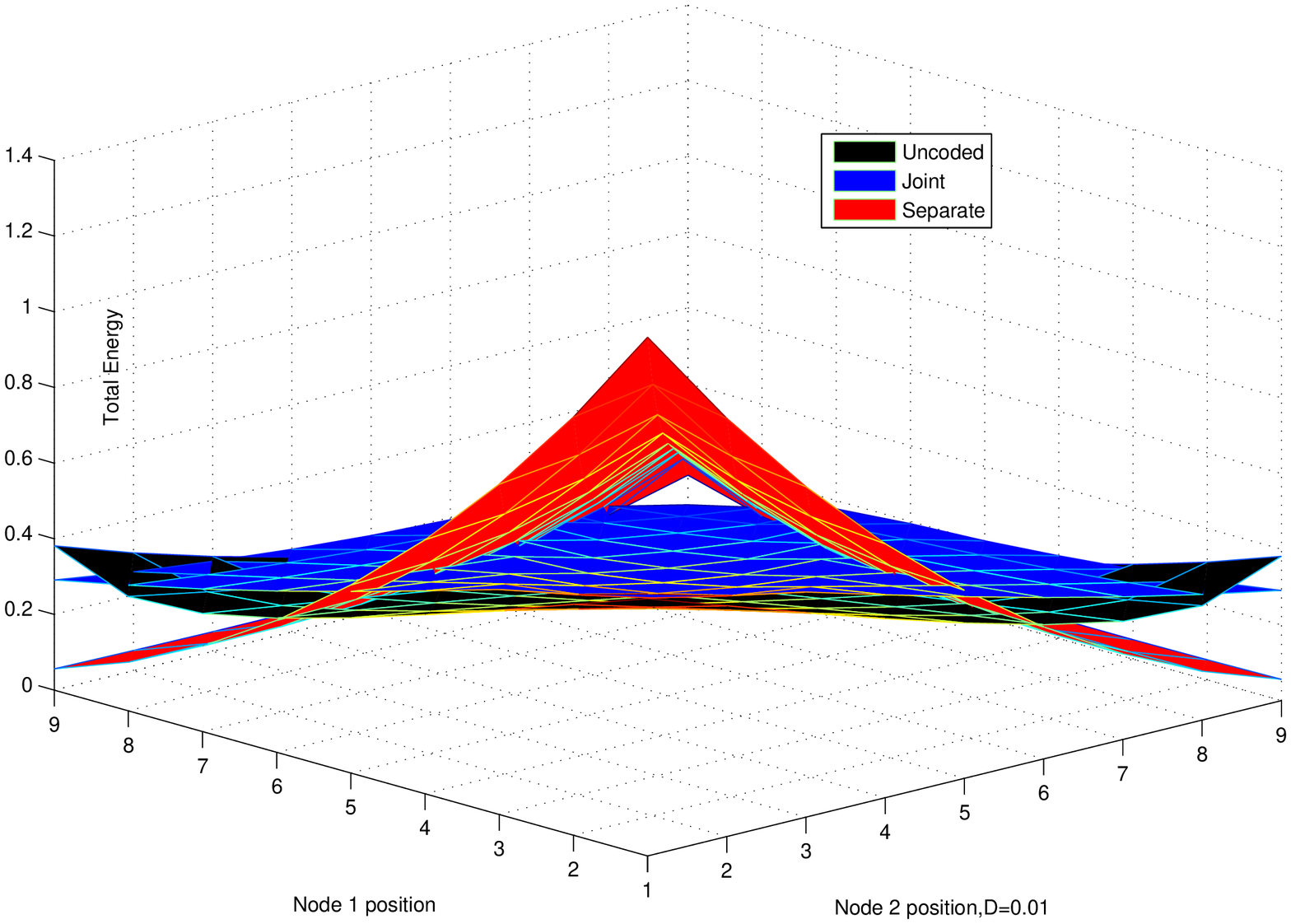,width=9.5cm}}
  \centerline{(c) Results 3: $D=0.01$, $\sigma_S^2 = \sigma_W^2=1$,
$\beta_c=\beta_s=2$}. \medskip
\end{minipage}
\caption{Total power consumption for separate source-channel
coding (Red), joint source-channel coding (Blue) and uncoded
(Black) schemes} \label{power results}
\end{figure}

We then run the geometrical programming based optimization algorithm to determine the minimum
total transmission powers for various approaches. We consider $9$
spots uniformly distributed between the source and fusion center
for possible locations of two sensors, which are indexed by
integers $1$ through $9$. The smaller the index value is, the
closer the sensor is located to the source.  Figure~\ref{power results}(a),
Figure~\ref{power results}(b) and Figure~\ref{power results}(c) demonstrate the
total minimum power
consumption $P_1+P_2$ as a function of nodes' locations for three 
sensor processing schemes, from which we have
following observations:

\begin{itemize}

\item As proved in Theorem~\ref{theorem5}, when network is symmetric,
$P_{total,A} < P_{total, J} < P_{total,S}$.

\item Under a relatively large distortion constraint (e.g. $D =0.5$),
uncoded scheme is the most energy efficient among the three candidates
for all sensor locations, as shown by Figure~\ref{power results}(a).

\item Under relatively small distortion constraints (e.g. $D=0.1$ and
$D=0.01$), separate coding approach becomes the most energy
efficient  when  the relative position of two sensors becomes more asymmetric.
For example, in both Figure~\ref{power results}(b) and Figure~\ref{power
results}(c), at a location with an index
pair $(1,9)$, i.e. the first sensor is closest to the source and the second
sensor is closest to the fusion center,  we have
$P_{total,A} > P_{total, J} > P_{total,S}$.

\item Overall, to minimize the total power expenditure, we should
choose either uncoded transmission or separate coding scheme for a
given pair of locations. This is a bit surprising as joint coded
approach is often advocated more efficient (rate wise) than the
separate one. It thus exemplifies that exact values of channel
conditions and
 the level of measurement noise are crucial  to concluding which scheme is the
 most power efficient in non-symmetric Gaussian networks with a finite number
of sensors.

\end{itemize}

\appendix

\section{Proof of Theorem~\ref{theorem correlated}}
\label{appendix1}
\begin{proof}

From Theorem~\ref{theorem2}, we know that each sensor finds from $2^{n \tilde{R_j}}$
codewords the closest one $U_j^{(k)}= \{U^{(k)}_j[i]\}$ to the observation vector $\{X_j[i]\}$ and then amplify-and-forwards
$\{Y_j[i]\}$ to the fusion center. The decoder  applies  jointly typical sequence decoding
\cite{cover_book} to seek $\{ U^{(k)}_j, j= 1, \cdots, L \}$ from $L$ codebooks which are jointly typical
with the received vector $\{Z[i]\}$.

WLOG, re-shuffle $2^{n \tilde{R}_j}$ vectors such that $U_j^{(1)}$  is the vector selected
by sensor $j \in \{1, \cdots, L\}$.
We assume that a subset $U^{(1)}(S^c) = \{ U^{(1)}_j, j \in S^c\}$ has been decoded correctly, while its complementary set
$U^{(1)}(S)= \{ U^{(1)}_j, j \in S\}$ is in error, which implies that
the channel decoder at fusion center is in favor of a set of vectors  $U^{(k)}(S) = \{ U^{(k_j)}_j, k_j \neq 1,  j
\in S\}$, instead.  Next, We  will find the upper bound of the  probability that $\left(U^{(1)}(S^c), Z
\right)$ and $U^{(k)}(S)$ are jointly typical.

The technique to upper-bond this probability is quite similar as the one for MAC channels with independent
channel inputs \cite[Chap 15.3]{cover_book}. The major difference here is that the channel inputs from $L$
sensors are correlated because of the testing channel model used in independent source coding, i.e. $U_j = X_j
+ V_j = X_0 + N_j + V_j$, for $j=1, \cdots, L$. 

The upper bound of the  probability that $\left(U^{(1)}(S^c), Z
\right)$ and $U^{(k)}(S)$ are jointly typical
is therefore
\begin{gather}
2^{n \left( H(U(S), U(S^c), Z)+\epsilon\right)} 2^{-n \left(H(U(S^c), Z)-\epsilon \right)} 2^{-n \sum_{i \in
S} \left( H(U_i) - \epsilon \right)} \label{eq:b1} \\
= \exp_2 \left(- n \left(H(U(S^c), Z) + \sum_{i \in S} H(U_i) \right. \right. \nonumber \\ \left. \left. - H(U(S), U(S^c), Z)-
(|S|+2)\epsilon \right)  \right) \label{eq:b2}
\end{gather}

\no where the first term in (\ref{eq:b1}) is the upper bound for the number of jointly typical sequences of
$(U(S), U(S^c), Z)$,
the second term in (\ref{eq:b1}) is the upper bound of the probability $P\left(U^{(1)}(S^c), Z\right)$
and the   last  term in (\ref{eq:b1}) is the upper bound of the probability
$P\left(U^{(k)}(S)\right)$. The summation in the last term in (\ref{eq:b1}) is due to the independence of
codebooks generated by each sensor and the assumption that decoder is in favor of some $U^{(k_j)}_j$ for
$k_j \neq 1$ and $j \in S$, which are independent of $U^{(1)}(S)$.

Since we have at most  $2^{n \sum_{j \in S} \tilde{R}_j}$  number of sequences to be confused with
$U^{(1)}_j, j \in S$, we need
\begin{gather} \label{eq:b3}
\sum_{j \in S} \tilde{R}_j < H(U(S^c), Z)\nonumber \\
 + \sum_{i \in S} H(U_i) - H(U(S), U(S^c), Z)-
(|S|+2)\epsilon \nonumber \\
= \sum_{i =1} ^{|S|-1} I\left(U_{\pi_i};
U_{\pi_{i+1}}^
{\pi_{|S|}} \right)  + I\left(U(S); U(S^c), Z\right) -
(|S|+2)\epsilon \end{gather}

\no for all  $S \subseteq \{ 1, \cdots, L \}$ and any arbitrarily small $\epsilon$
in order to  achieve the asymptotic zero error probability as $n \rightarrow \infty$,  which thus completes the proof.

\end{proof}

\section{Proof of Lemma~\ref{lemma mutual0}}
\label{appendix2}
\begin{proof}

As $U_2 \rightarrow X_2 \rightarrow
X_1 \rightarrow U_1$ forms a Markov chain, we have
\begin{gather}
I( U_2; X_2, X_1|U_1) \stackrel{(a)}{=} I(U_2; X_1, X_2, U_1) -
I(U_2;U_1)
\nonumber \\
 \stackrel{(b)}{=}  I(U_2; X_2) + I(U_2; X_1, U_1 | X_2) - I(U_2; U_1) \nonumber
\\
\stackrel{(c)}{=}  I(U_2; X_2) - I(U_2; U_1) \label{eq1}
\end{gather}

\no where equations $(a)$ and $(b) $ are due to the chain rule on
conditional mutual information \cite{cover_book}. Given Markov
chain of  $U_2 \rightarrow X_2 \rightarrow X_1 \rightarrow U_1$,
$U_2$ and $(X_1,U_1)$ are conditionally independent given $X_2$
and consequently $I(U_2; X_1, U_1 | X_2) =0$ leading to equation
$(c)$.

On the other hand,  following equations also hold under similar
arguments:
\begin{gather}
I( U_2; X_2, X_1|U_1) \stackrel{(1)}{=} I(U_2; X_2 |U_1) + I(U_2;
X_1 | X_2,
U_1) \nonumber \\
 \stackrel{(2)}{=} I(U_2; X_2 |U_1) + I(U_2; X_1, U_1|X_2) - I(U_2; U_1
|X_2)\nonumber \\
 \stackrel{(3)}{=}  I(U_2; X_2 |U_1) \label{eq2}
\end{gather}

\no Therefore, combining (\ref{eq1}) and (\ref{eq2}) yields: \beeq
\label{mutual I6} I(U_2; X_2|U_1) = I(U_2; X_2) - I(U_1; U_2)
\eeeq \no and similarly, \beeq I(U_1; X_1|U_2) = I(U_1; X_1) -
I(U_1; U_2) \eeeq

\no which thus proves (\ref{mutual I1}) and (\ref{mutual I2}).

We can prove (\ref{mutual I3}) by firstly showing  that \beeq
\label{mutual I4} I(U_1,U_2; X_1, X_2) = I(X_1, X_2; U_1) + I(U_2;
X_2,X_1|U_1) \eeeq

\no under the chain rule, where  \beeq I(X_1, X_2; U_1) =  I(X_1;
U_1) + I(X_2; U_1|X_1) = I(X_1; U_1) \eeeq

\no because of $U_2 \rightarrow X_2 \rightarrow X_1 \rightarrow
U_1$. Since we have already proved (\ref{eq1}), it is
straightforward to show that (\ref{mutual I3}) holds.

\end{proof}

\section{Proof of Lemma~\ref{lemma mutual3}}

\begin{proof}
Due to the independence of measurement noise, $U(S) \rightarrow X(S) \rightarrow X(S^c)
\rightarrow U(S^c)$ forms a  Markov chain. The first equation in (\ref{mutual I5}) is a
direct application of (\ref{mutual I1}).

The proof of the second equation is based upon another Markov
chain by adding the source random variable $X_0$ into the former
one: $U(S) \rightarrow X(S) \rightarrow X_0  \rightarrow X(S^c)
\rightarrow U(S^c)$. From this Markov chain, we can  deduce
\begin{gather}
I \left[ U(S); X_0 | U(S^c) \right] = I\left[ U(S^c), X_0;
U(S)\right] - I\left[ U(S^c); U(S)\right] \nonumber \\
= I\left[ X_0; U(S) \right] + I\left[ U(S^c); U(S) |X_0\right] -
I\left[ U(S); U(S^c)\right] \nonumber \\
 =I\left[ X_0; U(S) \right] - I\left[ U(S); U(S^c)\right]
 \label{mutual I8}
 \end{gather}

\no and
\begin{gather}
\sum_{i \in S} I\left[ U_i; X_i | X_0 \right] = I \left[ U(S);
X(S) |X_0 \right] \nonumber \\
 = I\left[ U(S); X(S) \right] + I \left[ U(S); X_0 | X(S) \right]
 - I\left[ X_0; U(S) \right] \nonumber \\
 = I\left[ U(S) ; X(S) \right] - I\left[ X_0; U(S) \right]
 \label{mutual I9}
\end{gather}

\no where the first equality is because of the conditional independence of $(X_i, U_i)$
given $X_0$.

It can be seen that (\ref{mutual I8}) and (\ref{mutual I9}) yields
\begin{gather}
 I\left[U(S); X_0 | U(S^c)\right] + \sum_{i \in S} I\left[ U_i;
X_i | X_0 \right] \nonumber \\ = I\left(U(S); X(S) \right) -
I\left(U(S); U(S^c)\right),
\end{gather}

\no which completes the proof for Lemma~\ref{lemma mutual3}.

\end{proof}

\section{Proof of Theorem~\ref{theorem4}}
\label{JSCC order}
\begin{figure}[ht]
  \centering
         \centerline{
        \scalebox{1.2}
          {
        \input{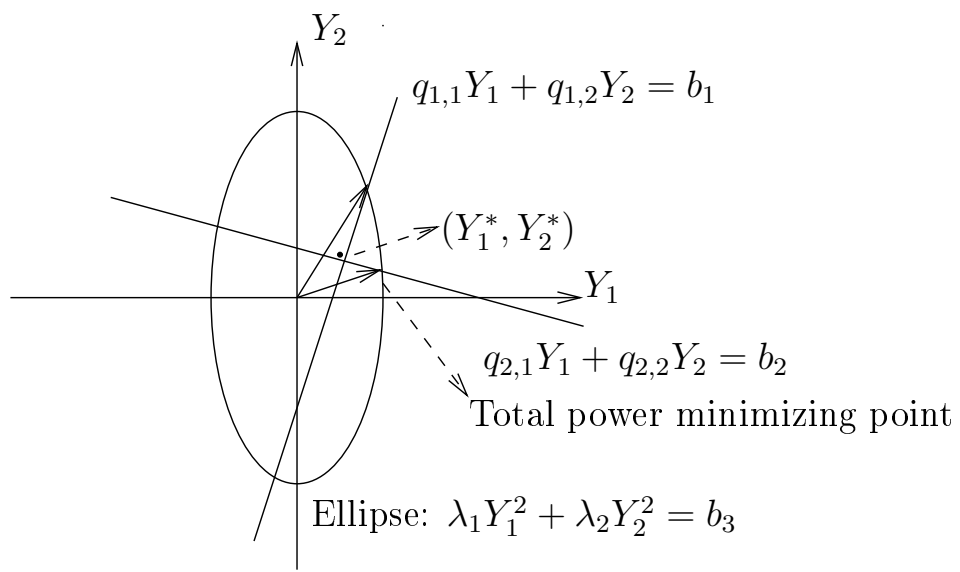}
      }
 }
    \caption{Total power minimization}
 \label{ellipse}
\end{figure}

\begin{proof}

Define $Z_i = 2^{2\tilde{R}_i} \left( 1- \tilde{\rho}^2 \right)$,
for $ i=1,2$. It can be shown that $Z_i >1$ by using
$\tilde{\rho}^2 = \rho^2 ( 1- 2^{-2 \tilde{R}_1}) ( 1- 2^{-2
\tilde{R}_2})$ and $0< \rho <1$ as shown in (\ref{direct rho}). 

Given a pair of quantization rates $(\tilde{R}_1, \tilde{R}_2)$,
the original optimization problem becomes
\begin{gather}
\min P_1 + P_2, \; \mbox{subject to:} \nonumber \\
P_1 \geq \frac{(Z_1 -1)\sigma_W^2}{(1- \tilde{\rho}^2)
g_1}\stackrel{\Delta}{=}b_1, \; P_2 \geq \frac{(Z_2
-1)\sigma_W^2}{(1- \tilde{\rho}^2) g_2} \stackrel{\Delta}{=}b_2
\nonumber \\
P_1 g_1 + P_2 g_2 + 2 \tilde{\rho}  \sqrt{P_1 g_1 P_2 g_2}
\nonumber \\
\geq \sigma_W^2\left( 2^{2\tilde{R}_1+2 \tilde{R}_2}( 1-
\tilde{\rho}^2 ) -1  \right) \stackrel{\Delta}{=}b_3 \label{new
opt}
\end{gather}

Define a matrix \beeq {\bf A} = \left[ \begin{array}{ll} g_1 &
\tilde{\rho} \sqrt{g_1 g_2} \\
\tilde{\rho} \sqrt{g_1 g_2} & g_2 \end{array}\right] \eeeq

\no whose  eigenvalue decomposition is:  ${\bf A} = {\bf Q}
\Lambda {\bf Q}^{T}$, where the diagonal matrix $\Lambda =
\mbox{diag}\{\lambda_1, \lambda_2\}$ has eigenvalues $\lambda_i$
of ${\bf A}$ and the column vectors of the matrix  \beeq {\bf Q} =
\left[
\begin{array}{ll} q_{1,1} & q_{1,2} \\
q_{2,1} & q_{2,2} \end{array}\right] \eeeq

 \no are normalized eigenvectors associated with $\lambda_1$ and $\lambda_2$
 respectively, which satisfy:
 \beeq
{\bf Q}^T {\bf Q} =  \left[ \begin{array}{ll} 1 & 0 \\ 0 & 1
\end{array}\right] \stackrel{\Delta}{=} {\bf I_2}
 \eeeq

Notice that  the last constraint in (\ref{new opt}) is a quadratic
form of variables $\sqrt{P_1}$ and $\sqrt{P_2}$, which essentially
determines an ellipse, we can perform an unitary transformation by
introducing two new variables $Y_1$ and $Y_2$: \beeq {\bf Y} = [
Y_1, Y_2]^T = {\bf Q}^T [\sqrt{P_1}, \sqrt{P_2}]^T\eeeq

\no such that the original optimization problem in (\ref{new opt})
is transformed to an equivalent one:
\begin{gather}
\min ||Y||^2 = Y_1^2 + Y_2^2, \; \mbox{subject to:} \nonumber \\
{\bf Q}[Y_1, Y_2]^T \geq [b_1, b_2] \nonumber \\
\lambda_1 Y_1^2 + \lambda_2 Y_2^2 \geq b_3. \label{ellipse opt}
\end{gather}

\no
We show next that  $ Y_1^2 + Y_2^2$ is minimized  at the point
where the second line $q_{2,1} Y_1 + q_{2,2} Y_2 = b_2$ intersects
with the ellipse $\lambda_1 Y_1^2 + \lambda_2 Y_2^2 = b_3$ as shown in Figure~\ref{ellipse}. In order to prove
this, we need to first prove that the intersection of two lines $q_{i,1} Y_1 +
q_{i,2} Y_2 = b_i, i=1,2$ is inside the ellipse. Let $[Y_1^*, Y_2^*]$ denote
the crossing point of the two lines, which can be determined as
$[Y_1^*, Y_2^*] = {\bf Q}^T [b_1, b_2]^T$. It is sufficient to prove that
$\lambda_1 (Y_1^*)^2 + \lambda_2 (Y_2^*)^2 < b_3$, which is equivalent to
having
\beeq
[b_1, b_2] {\bf A} [b_1, b_2]^T < b_3.
\eeeq

\no This holds as
\begin{gather}
[b_1, b_2] {\bf A} [b_1, b_2]^T- b_3 \nonumber \\
 = -\frac{\sigma_W^2}{1-
\tilde{\rho}^2}
\left[ \sqrt{(Z_1-1) (Z_2-1)} -\tilde{\rho}\right]^2 <0
\end{gather}

\no where $Z_i$ were defined right below (\ref{eq: theorem4}), and hence obtain
the desired result.

Assume $\lambda_1> \lambda_2$. Solving quadratic function of
$|\lambda {\bf I_2} - {\bf A}|=0$, we obtain eigenvalues
$\lambda_1$ and $\lambda_2$: \beeq \label{lambda} \lambda_{1,2} =
\frac{1}{2} (g_1+g_2) \left[ 1 \pm \sqrt{1 - \Delta} \right] \eeeq

\no where $\Delta = \frac{4 g_1 g_2 ( 1-
\tilde{\rho}^2)}{(g_1+g_2)^2}$. The entries of eigenvectors can be
computed  accordingly:
\begin{gather}
q_{1,1} = \sqrt{\frac{\lambda_1 - g_2}{2 \lambda_1 - g_1 -g_2}},\;
q_{1,2} = -\sqrt{\frac{\lambda_2 - g_2}{2 \lambda_2 - g_1 -g_2}}
\nonumber \\
q_{2,1} = \sqrt{\frac{\lambda_1 - g_1}{2 \lambda_1 - g_1 -g_2}},\;
q_{2,2} = \sqrt{\frac{\lambda_2- g_1}{2 \lambda_1 - g_1
-g_2}}.\label{eivector}
\end{gather}

Based on (\ref{lambda}), we have $2\lambda_1 - g_1 - g_2 = -(2
\lambda_2 - g_1 - g_2) = (g_1+g_2) \sqrt{1-\Delta}$. Due to the
non-negativeness of the ratios involved in (\ref{eivector}), it
can be shown that $\lambda_1 > g_1> g_2 > \lambda_2$. In addition,
because of $\lambda_1 + \lambda_2 = g_1+g_2$, $q_{i,j}$'s satisfy:
\beeq q_{1,1} > |q_{1,2}|, \; q_{2,1} < q_{2,2}. \eeeq

\no
We can therefore conclude that the lengths of the semi-axis
$1/\sqrt{\lambda_1}$ and $1/\sqrt{\lambda_2}$ of the ellipse in
the direction of $Y_1$ and $Y_2$, respectively, satisfy $
1/\sqrt{\lambda_1}< 1/\sqrt{\lambda_2}$.

Also, since the slopes of
the lines $q_{2,1} Y_1 + q_{2,2} Y_2 = b_2$  and $q_{1,1} Y_1 +
q_{1,2} Y_2 = b_1$ have the relationship of $q_{2,1}/q_{2,2}< 1<
q_{1,1}/|q_{1,2}|$, in addition, $[Y_1^*, Y_2^*]$ is inside the ellipse, the
minimum distance in (\ref{ellipse opt}) is
attained at the point where the line with smaller slope intersects
with the ellipse, as illustrated by Figure~\ref{ellipse}, which
  implies to minimize the total
transmission
power $P_1 + P_2$, the second constraint on $P_2$ in (\ref{new
opt}), as well as the third one, should be active. This is
equivalent to having: $\tilde{R}_2 = I(U_2; Z, U_1)$ and
$\tilde{R}_1 + \tilde{R_2} = I(U_1, U_2; Z) + I(U_2; U_1)$, and
$\tilde{R}_1 = I(U_1;Z)$.

Consequently, when $g_1>g_2$, the decoding at the fusion center
follows exactly as that described in Theorem~\ref{theorem4}.

\end{proof}

\bibliographystyle{IEEEbib}
\bibliography{IPSN2007_bib}

\end{document}